\documentclass[aps,prl,twocolumn,superscriptaddress,groupedaddress]{revtex4}  % for review and submission

\usepackage{graphicx}  % needed for figures
\usepackage{dcolumn}   % needed for some tables
\usepackage{bm}        % for math
\usepackage{amssymb}   % for math
\usepackage{lingmacros}
\usepackage{tree-dvips}
\usepackage{mathtools}
\usepackage[utf8]{inputenc}
\usepackage{graphicx}
\usepackage[rightcaption]{sidecap}
\usepackage{float}
\usepackage{wrapfig}
\usepackage{amsmath}
\usepackage{parskip}
\usepackage{atbegshi}
\usepackage{pdfpages}
\usepackage{xcolor}
\usepackage{siunitx}
\DeclareSIUnit{\density}{\kilogram\per\cubic\metre}
\DeclareSIUnit{\tension}{\newton\per\metre}

\usepackage{titlesec}
\titlespacing*{\section}{0pt}{1\baselineskip}{\baselineskip}
\titleformat*{\section}{\large\bfseries}
\usepackage{color}

\begin{document}

\title{Characterizing multi-mode nonlinear dynamics of nanomechanical resonators}

	\author{Ata Keşkekler\footnote{Corresponding authors: \\ Ata Ke\c{s}kekler $<$a.keskekler-1@tudelft.nl$>$, \\ Farbod Alijani $<$f.alijani@tudelft.nl$>$}}
	% \affiliation{Department of Precision and Microsystems Engineering, TU Delft, The Netherlands}
	\affiliation{Faculty of Mechanical, Maritime and Materials Engineering, Delft University of Technology, Mekelweg 2, 2628 CD Delft, The Netherlands}
 
	\author{Vincent Bos}
	% \affiliation{Department of Precision and Microsystems Engineering, TU Delft, The Netherlands}
	\affiliation{Faculty of Mechanical, Maritime and Materials Engineering, Delft University of Technology, Mekelweg 2, 2628 CD Delft, The Netherlands}
	
	\author{Alejandro M. Aragón}
	% \affiliation{Department of Precision and Microsystems Engineering, TU Delft, The Netherlands}
	\affiliation{Faculty of Mechanical, Maritime and Materials Engineering, Delft University of Technology, Mekelweg 2, 2628 CD Delft, The Netherlands}
	
	\author{Peter G. Steeneken}
	\affiliation{Faculty of Mechanical, Maritime and Materials Engineering, Delft University of Technology, Mekelweg 2, 2628 CD Delft, The Netherlands}
	\affiliation{Kavli Institute of Nanoscience, TU Delft, The Netherlands}

	\author{Farbod Alijani$^\ast$}
	% \affiliation{Department of Precision and Microsystems Engineering, TU Delft, The Netherlands}
	\affiliation{Faculty of Mechanical, Maritime and Materials Engineering, Delft University of Technology, Mekelweg 2, 2628 CD Delft, The Netherlands}
	
	% D0 authors (remove the first 3 lines
	% of this file prior to submission, they
	% contain a time stamp for the authorlist)
	% (includes institutions and visitors)
	\date{\today}
	\begin{abstract}
	    
     Mechanical nonlinearities dominate the motion of nanoresonators already at relatively small oscillation amplitudes. Although single and coupled two-degrees-of-freedom models have been used to account for experimentally observed nonlinear effects, it is shown that these models quickly deviate from experimental findings when multiple modes influence the nonlinear response. Here, we present a nonlinear reduced-order modelling methodology based on FEM simulations for capturing the global nonlinear dynamics of nanomechanical resonators. Our physics-based approach obtains the quadratic and cubic nonlinearities of resonators over a wide frequency range that spans \SI{70}{\mega\hertz}. To qualitatively validate our approach, we perform experiments on a graphene nanodrum driven opto-thermally and show that the model can replicate diverse ranges of nonlinear phenomena, including multi-stability, parametric resonance, and different internal resonances without considering any empirical nonlinear fitting parameters. By providing a direct link between microscopic geometry, material parameters, and nonlinear dynamic response, we clarify the physical significance of nonlinear parameters that are obtained from fitting the dynamics of nanomechanical systems, and provide a route for designing devices with desired nonlinear behaviour.  
	    \end{abstract}

	\maketitle

Nanomechanical resonators are the devices of choice for high-performance sensing since they respond to minuscule forces \cite{Chaste2012,Fogliano2021,roslon2022probing}.
More recently, they have emerged as ideal systems for exploring nonlinear dynamic phenomena. Thanks to their high force sensitivity they are easily driven into the nonlinear regime \cite {bachtold2022mesoscopic}. Their small mass leads to high resonance frequencies, which facilitates high-speed measurements and, especially in ultra-thin resonators, the high aspect ratio allows tuning of tension and resonance frequencies to explore a variety of nonlinear phenomena \cite{Steeneken2021}. 

When nanomechancial devices are driven into resonance, already at small amplitudes Duffing nonlinearities precipitate in the motion, leading to softening- or hardening-type nonlinear responses~\cite{lifshitzcross}. When the nonlinear regime is traversed further, with the increase in drive amplitude, other eigenmodes begin to partake in the motion through autoparametric excitations and nonlinear intermodal couplings \cite{NayfehOscillations}. In this nonlinear domain, many studies have reported significant impact of these nonlinear couplings on the effective dissipation and stiffness of the vibration modes~\cite{Guttinger2017,Keskekler2021}. A number of exotic nonlinear phenomena have also been showcased, ranging from frequency noise suppression~\cite{antonio2012frequency} and intermodal storage of mechanical energy~\cite{periodtriple,coherentEnergy}, to the generation of mechanical frequency combs~\cite{Keskekler2022,oriComb}.

Efforts made to date in explaining nonlinear phenomena often rely on proposing analytical nonlinear low-degrees-of-freedom (DOFs) models that are fit to the experimental data to prove their validity. However, since there is no direct link between the magnitude of the resulting fit parameters and the geometry or material properties of the nanomechanical device, it is difficult to evaluate whether these models are the only ones that can account for the experimental data, and which parameters are the most relevant. Moreover, since there is no direct link between the model parameters and the underlying physics, it is difficult to extract device information from the fitting. A realistic description of the complex nonlinear dynamics of nanoresonators with pure analytical methods such as rotating-wave approximation~\cite{markBSpert} or harmonic balancing~\cite{dejan} is not always sufficiently accurate, because analytical methods are constrained to a limited number of DOFs. Purely numerical methods such as molecular dynamics or dynamic nonlinear Finite Element Method (FEM) simulations may resolve this issue, yet they are computationally expensive~\cite{FPUgraph,SAJADI} and provide less insight. Therefore, an intermediate approach, whereby analytical multi-mode nonlinear dynamic models are constructed from the physical device properties using numerical methods, can be extremely valuable for the precise and fast analysis of the nonlinear dynamics of nanomechanical systems.

\begin{figure*}
    \centering
    \includegraphics[width=1\linewidth]{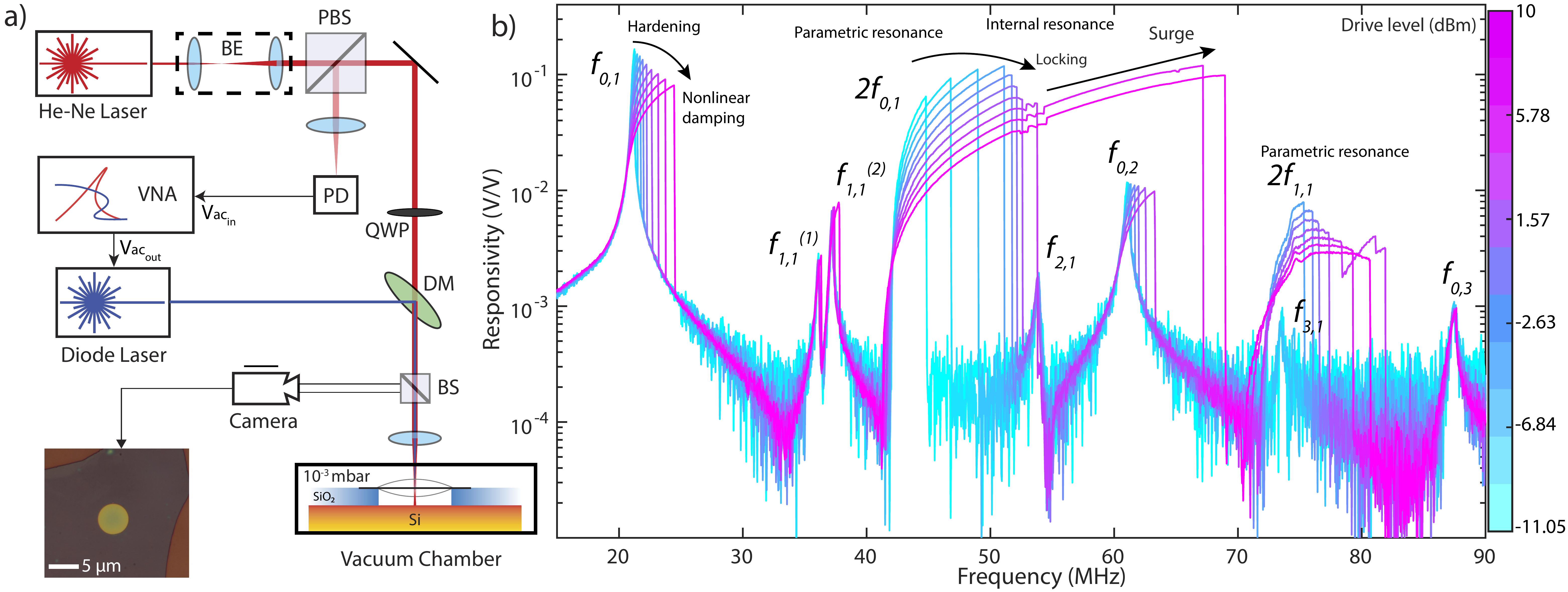}
    \caption{Measuring the motion of a nonlinear graphene nanodrum. (a) Schematic of the measurement setup. BE, PBS, BS, QWP, DM and VNA stand for beam expander, polarized beam splitter, beam splitter, quarter-wave plate, dichroic mirror and vector network analyzer, respectively. (b) Measurements reveal a range of nonlinear dynamic phenomena for the graphene nanodrum at relatively small amplitudes. At high drive levels, graphene nanodrum exhibits complex nonlinear behavior where the frequency response shows hardening nonlinearity, signatures of nonlinear damping, parametric resonance, mode couplings and internal resonances. The labels $f_\text{p,m}$ are frequencies associated with circular drum mode shapes that are found from FEM simulations. Here, $p$ stands for the number of nodal lines and $m$ is the number of nodal circles.}
    \label{fig:Fig1}
\end{figure*}

	In this article we develop and utilize a physics-based Reduced-Order Model (ROM) to characterize multi-mode nonlinear dynamics of nanomechanical resonators over a wide frequency range. Our approach makes use of FEM simulations to probe the geometric nonlinearities of nanoresonators for a large number of coupled vibrational modes. To validate our method, we perform experiments on a graphene nanodrum that is driven opto-thermally into the strong mode coupling regime. We show that the physics-based model can capture the response of the seven directly excited and two parametrically excited modes of the graphene nanodrum from linear to nonlinear regime, in a frequency range that spans \SI{70}{\mega\hertz}. The model uses the Young’s modulus and pre-stress to obtain the coupling coefficients, thus providing insight into the influence of geometric and physical parameters on the coupled dynamics of nanomechanical resonators. Since the method is FEM-based, it can be applied further to nanomechanical devices of virtually any geometry, allowing predicting and designing a variety of nonlinear phenomena. 
	
	As an experimental model system for demonstrating the method, we probe the complex dynamics of a graphene nanodrum resonator. The resonator is fabricated by dry transfer of $h=\SI{10}{\nano\metre}$ thick multi-layer graphene over a $d=\SI{5}{\micro\metre}$ diameter and \SI{285}{\nano\metre} deep circular cavity, etched in a layer of SiO$_2$ on a Si substrate. To study the mechanical vibrations of the nanodrum, we opto-thermally drive it using a blue laser ($\lambda=\SI{405}{\nano\metre}$) and measure its response by a red laser ($\lambda=\SI{633}{\nano\metre}$) using laser interferometry (Fig.~\ref{fig:Fig1}a). At low drive powers, a linear set of resonance peaks can be obtained, showing the activation of multiple modes of vibration that we can identify easily using FEM simulations. As the drive level is increased, the nanoresonator quickly shows signs of nonlinearity (Fig.~\ref{fig:Fig1}b). It is possible to observe the well-studied Duffing (hardening type) nonlinearity in several modes already at relatively small amplitudes (below \SI{-6}{\decibel m}).

 By further increasing the drive level, we notice rapid activation of a plethora of nonlinear dynamic responses. For instance, when the excitation frequency is tuned to twice the resonance frequency of the modes $f_{0,1}$ and $f_{1,1}$, it is possible to detect strong parametric resonances~\cite{dolleman2018opto}. Since the tension of the nanodrum is directly related to its stiffness, modulation of the tension via opto-thermal actuation parametrically excites the nanodrum. Consequently, for conditions where the drive is strong enough, period doubling instabilities emerge, resulting in parametric resonances~\cite{dolleman2018opto}. These resonances can reach high amplitudes and span wide frequency ranges thanks to the Duffing hardening nonlinearity. Especially at drive frequencies where the frequencies of these modes satisfy internal resonance conditions~\cite{NayfehOscillations}, they strongly interact with other modes of vibration. This can be observed in Fig.~\ref{fig:Fig1}b at drive levels above \SI{2}{\decibel m}, around the region where the parametrically driven $f_{0,1}$ and the directly driven $f_{2,1}$ are interacting. As the parametric response of $f_{0,1}$ approaches  $f_{2,1}$, it is also possible to observe a  decrease in the responsivity as well as a reduction in the rate of increase in the nonlinear frequency of the parametric resonance -- a phenomenon that we label as ``locking'' in Fig.~\ref{fig:Fig1}b. Only after a certain drive level is reached, this ``locking'' barrier is surpassed, and the parametric resonance surges to a higher amplitude and frequency~\cite{Keskekler2021}. Other than this apparent interaction, a similar coupled dynamic response can be  noticed in the neighborhood of the parametrically driven $f_{1,1}$ and directly driven $f_{3,1}$.

	Another interesting observation is the decrease in the responsivity of the nanoresonator for the directly driven modes of vibration with the increase in drive level (see the nonlinear response around $f_{0,1}$ and $f_{0,2}$ in Fig.~\ref{fig:Fig1}b for instance). This reduction, which is a result of the emergence of nonlinear dissipation~\cite{lifshitzcross}, was first observed in resonators that can reach strong nonlinear regime~\cite{eichler2011}. It was recently shown that such nonlinear dissipation process could also be mediated when two modes of vibration are coupled via internal resonance~\cite{Keskekler2021}. Capturing such nonlinear dissipation processes by analytical means, however, when multiple modes are contributing to the response, is far from being trivial. In order to deepen our understanding of these complex multi-modal interactions, we thus introduce a FEM-based ROM model~\cite{MURAVYOV20031513}.
 
\begin{figure}
    \centering
    \includegraphics[width=1\linewidth]{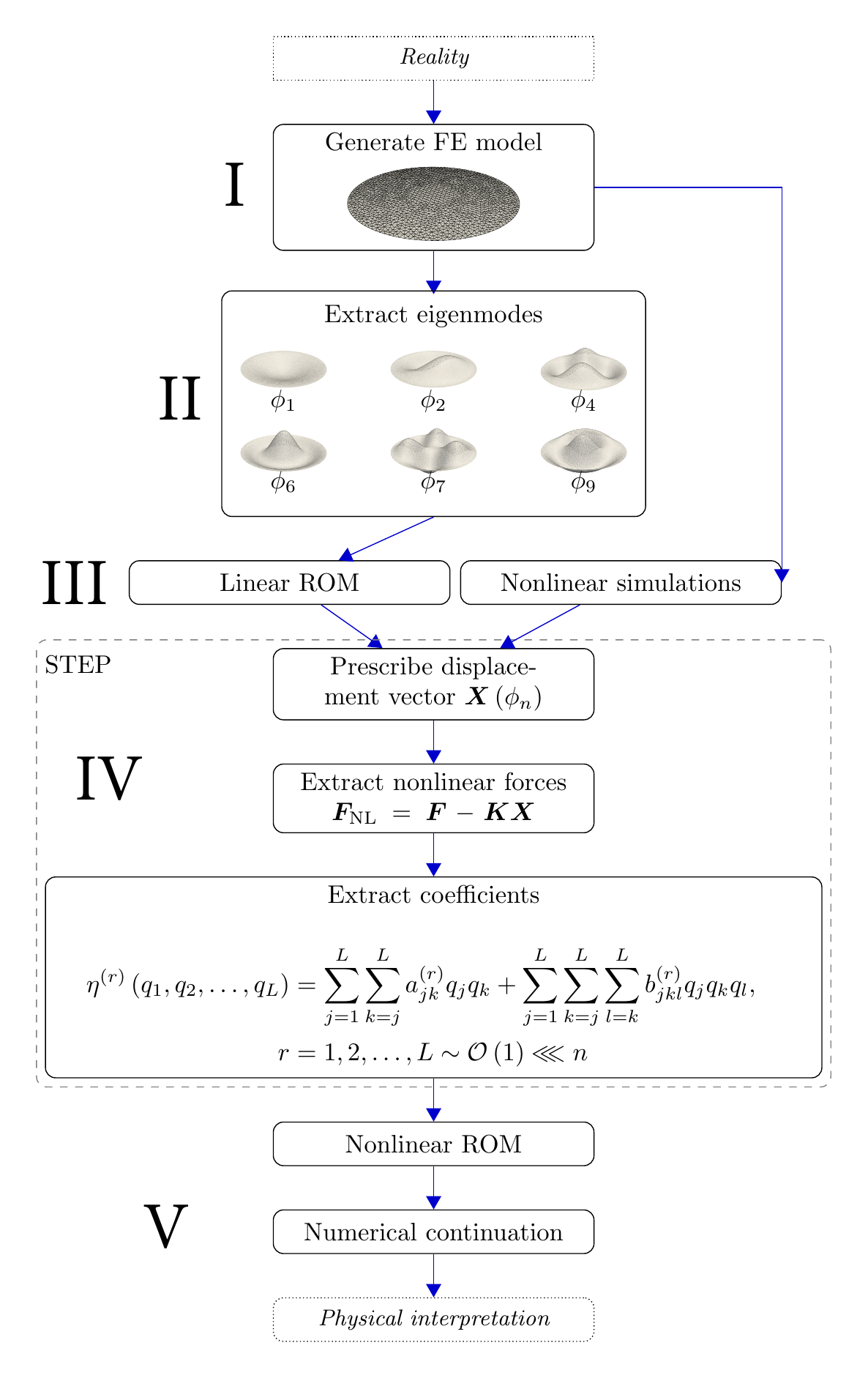}
    \caption{The flowchart of the ROM procedure for nonlinear dynamic simulations. The frequency response and AFM measurements are used to extract the physical parameters for building the FE model, which is then utilized to obtain linear and nonlinear reaction forces of the device, given prescribed displacements in terms of superposition of eigenmodes. These forces are then used for extracting the coefficients of nonlinear terms in the ROM \cite{MURAVYOV20031513}. Finally the nonlinear multi-mode model is simulated numerically to obtain the full nonlinear dynamic response.}
    \label{fig:fig2}
\end{figure}
\begin{figure*}
    \centering
    \includegraphics[width=1\linewidth]{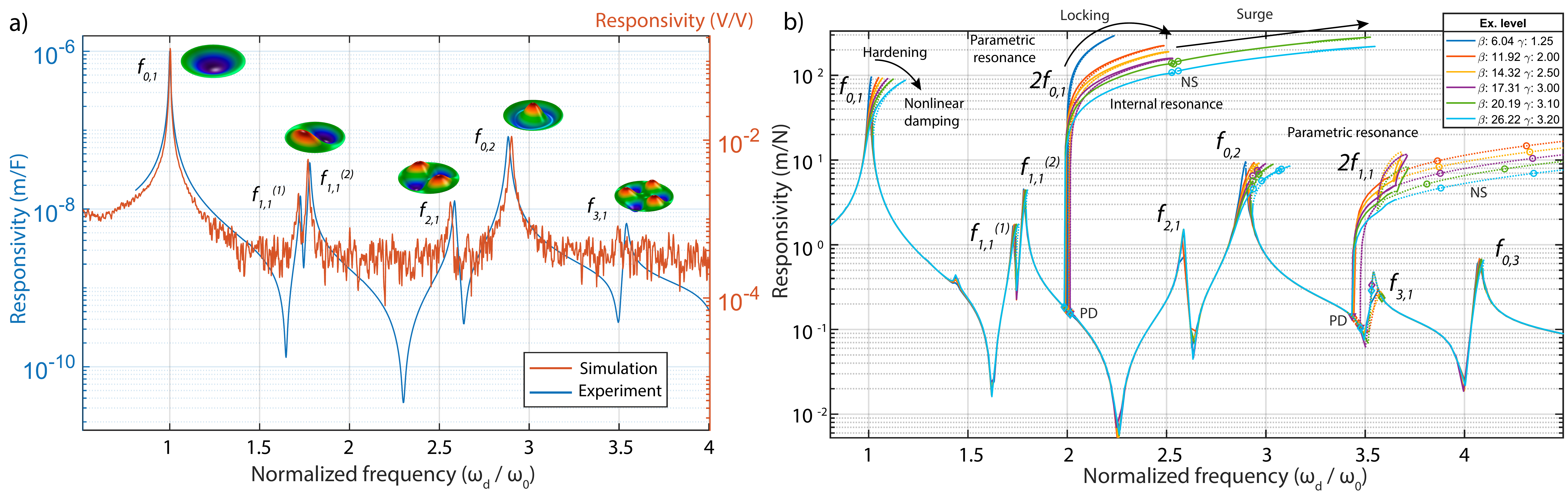}
    \caption{The simulated linear and nonlinear dynamic responses of the ROM. a) Frequency response measurements at linear regime are used for extracting the pre-tension and Young's Modulus of the membrane. b) The simulated nonlinear frequency response of the graphene nanodrum, where $\beta$ is the direct drive intensity and $\gamma$ is the parametric drive intensity. It is possible to capture the experimentally observed nonlinear phenomena using the ROM , such as hardening nonlinearities, parametric resonances and internal resonance induced physics such as frequency locking and amplitude surge. Bifurcation points are also detected during numerical  simulations, diamond indicators stand for period doubling bifurcations (PD), whereas circles stand for the Neimark-Sacker bifurcation (NS).}
    \label{fig:fig3}
\end{figure*}

The method for modelling the complex dynamics showcased in Fig.~\ref{fig:Fig1}b consists of five steps, starting with the generation of a FEM model of the nano device (step I in Fig.~\ref{fig:fig2}). For this purpose, any FEM software package that can handle geometric nonlinearities can be used (we have chosen to use COMSOL in this work). The FEM model of the circular graphene nanodrum resonator uses plate elements and fixed boundary condition. We use the literature values for the mass density and Poisson's ratio of graphene to be $\rho=\SI{2267}{\density}$ and $\nu=0.16$. As the geometry of the resonator is already known from optical microscopy and Atomic Force Microscopy (AFM) measurements, this leaves only two unknown parameters to be determined for building the  model, namely the pre-tension and the Young's modulus of the nanodrum. Since the linear resonance frequencies of the graphene membrane are dominated by the pre-tension, we can extract its value from frequency response measurements at low drive levels. However, if only the fundamental frequency is taken into account when determining the pre-tension, it is not possible to explain the splitting of the asymmetric modes, like $f_{1,1}^{(1)}$ and $f_{1,1}^{(2)}$ in Fig.~\ref{fig:fig3}a. In a perfectly symmetric drum, these eigenmodes are degenerate, i.e., they have the same frequency. But in practice, there is a mismatch in the tension along the in-plane axes that causes these modes to have slightly different frequencies. The frequencies of the first degenerate mode, together with the fundamental frequency, are enough to extract the tension in the membrane by matching the experimental linear resonance frequencies of the first three modes in the FEM analysis. By doing this we found the pre-tension in the membrane to be $T_x=\SI{0.321}{\tension}$ and $T_y=\SI{0.257}{\tension}$ for two perpendicular axes in the plane.
To find Young's modulus of the graphene nanodrum we then used the linear resonance frequencies of the higher modes following the method described in~\cite{SAJADIbending}; we found a Young's modulus of \SI{410}{\mega\pascal} (Fig.~\ref{fig:fig3}a), which is within the values reported in the literature~\cite{gomezmech2015,Isacsson_2017}.

By performing linear vibration analysis in the FEM software, we can now obtain the $n \times n$ linear mass and stiffness matrices ($\bm{K}$ and $\bm{M}$, respectively), as well as the eigenvalues $\omega_n$ and eigenmode matrix $\bm{\Phi}$,  where $n$ is the number of DOFs used in the FEM simulations (step II in Fig.~\ref{fig:fig2}). Therefore,  for an $n$-dimensional displacement vector $\bm{X}$ in the FEM model, we obtain the following set of discrete equations:
\begin{equation} \label{eq:discrete_system}
\bm{M}\bm{\Ddot{X}}(t) + \bm{C\Dot{X}}(t)+ \bm{KX}(t) + \bm{H}(\bm{X}(t))=\bm{F}(t),
\end{equation}
where $\bm{\Ddot{X}}$ denotes acceleration, $\bm{\Dot{X}}$ velocity, $\bm{H}$ the nonlinear force vector, and $\bm{F}(t)$ is the nodal force vector. Here, the linear damping matrix $\bm{C}$ accounts for dissipation. Currently nonlinear damping is not yet included in the equation of motion, although viscous material damping might be added via an imaginary term in the material's Young modulus. Evidently, in a finely meshed FEM model, the large number of degrees of freedom $n$ in Eq.~\eqref{eq:discrete_system} in combination with the wide frequency range, makes it practically impossible to use the FEM for simulating nonlinear dynamic responses like those in Fig.~\ref{fig:Fig1}b. Therefore, we use a subset $L$ of the $n$ eigenmodes for explaining the observed physics, where $L\ll n$. This mathematically means reducing the number of DOFs to only a few that are capable of replicating the nonlinear dynamics of the full model. To do so, we use the modal coordinate transformation $\bm{X}=\bm{\Phi} \bm{q}$, expressing the displacement as a superposition of eigenmode shapes, and only select a subset $\bm{\Phi}_{n\times L}$ of eigenvectors, such that $\bm{q}$ is the $L$-dimensional modal amplitude vector. Using this transformation, Eq.~\eqref{eq:discrete_system} can be re-written in modal coordinates as

\begin{equation} \label{eq:discrete_system_transformed}
\bm{\Tilde{M}}\bm{\Ddot{q}}(t) + \bm{\Tilde{C}\Dot{q}}(t)+ \bm{\Tilde{K}q}(t) +\bm{\eta}(\bm{q}(t))=\bm{\Tilde{F}}(t),
\end{equation}
where $\bm{\Tilde{M}}=\bm{\Phi}^T \bm{M} \bm{\Phi}$,  $\bm{\Tilde{C}}=\bm{\Phi}^T \bm{C} \bm{\Phi}$, $\bm{\Tilde{K}}=\bm{\Phi}^T \bm{K} \bm{\Phi}$, $\bm{\eta}=\bm{\Phi}^T \bm{H}$ and $\bm{\Tilde{F}}(t)=\bm{\Phi}^T \bm{F}(t)$.

From the linear FEM eigenmode simulation, all these matrices and vectors except $\bm{\eta}$ can be determined (step I-III in Fig.~\ref{fig:fig2}). To obtain this nonlinear matrix we perform multiple nonlinear stationary FEM simulations, with suitably chosen displacements $\bm{X}_c$ along the Stiffness Evaluation Procedure (STEP, see IV in Fig.~\ref{fig:fig2}) that we briefly outline in what follows and along the lines of Ref.~\onlinecite{MURAVYOV20031513}.

For any nodal displacement vector $\bm{X}=\bm{X}_c$, the reaction forces can be transformed into the modal domain and used for the extraction of nonlinear internal forces of the nanodrum. We do this by carefully prescribing nodal displacement vectors $\bm{X}_c$, to calculate the corresponding linear reaction forces $\bm{F_{\text{L}}}$ since $\bm{F_{\text{L}}} = \bm{K}\bm{X}_c$. After finding the linear reaction forces, we perform a full nonlinear static  analysis in the FEM package, considering that the nanodrum is subjected to the same displacement vector $\bm{X}_c$, and obtain the total nodal reaction force $\bm{F}_{\text{T}}$. By subtracting the linear reaction forces from this full static solution, i.e., $\bm{F_{\text{NL}}} = \bm{H}(\bm{X}_c)=\bm{F}_{\text{T}}(\bm{X}_c) - \bm{K}\bm{X}_c$, we then obtain the nonlinear reaction force and map that on the subset of eigenmodes selected as follows:  $\bm{\eta}=\bm{\Tilde{F}_{\text{NL}}}=\bm{\Phi}^T \bm{F_{\text{NL}}}$ (see steps III and IV in Fig.~\ref{fig:fig2}). We  finally expand this nonlinear reaction force for every mode of vibration in terms of quadratic and cubic nonlinear terms as follows:
\begin{equation}
\eta^{(r)}=\sum^L_{j=1}\sum^L_{k=j}\alpha^{(r)}_{jk}q_{j}q_{k}+\sum^L_{j=1}\sum^L_{k=j}\sum^L_{l=k}b^{(r)}_{jkl}q_{j}q_{k}q_{l},
\label{eq:Forces}
\end{equation}
where $r$ stands for the $r$th equation of motion also associated with the $r$th mode, and $j$, $k$, $l$ are the mode numbers. Furthermore, $L$ is the number of modes that is being considered in the ROM. We note that these cubic and quadratic nonlinear terms for single or two-mode models are commonly used to simulate nonlinear dynamics of nanomechanical resonators. However, here we extract them purely from geometric nonlinearities and can expand that to any number of modes of vibration, up to the number of DOFs that is being considered in the FEM simulations.

The procedure explained above enables us to generate a set of linearly independent $F_{\text{NL}}$ equations by applying different displacement vectors $\bm{X}_c(\phi_r), r=\left\{1,\ldots L\right\}$; these are selected such that they are superpositions of eigenvectors $\phi_r$ where each combination provides unique information about a nonlinear term through simulations, e.g., $X_c=\pm\bm{\phi_{j}} q_j \pm\bm{\phi_{k}} q_k \pm\bm{\phi_{l}} q_l$. In order to obtain information about the nonlinear reaction forces, the modal amplitudes in the displacement vector $\bm{X}_c$ should be chosen such that the nano device reaches the geometric nonlinear regime.\\
To clarify the procedure outlined above, we demonstrate how the nonlinear coefficients of the ROM can be derived for two hypothetical generalized coordinates, $q_1$ and $q_2$. We start by determining the uncoupled nonlinear coefficients of the system. For the first generalized coordinate $q_1$, we construct displacement vectors $\bm{X}_c$  from $\bm{\phi}_1$, such that only nonlinear terms associated with $q_1$ are activated:
\begin{align}
 \bm{X}_{1}=+\bm{\phi}_{1} q_1, \label{eq:disp_1}\\
\bm{X}_{2}=-\bm{\phi}_{1} q_1. \label{eq:disp_2}
\end{align}
This results in two equations with two unknowns:
\begin{align}
\bm{\Tilde{F}_{\text{NL}}}^{(1)}(\bm{X}_{1})=\alpha_{11}^{(1)} q_{1}^2+b_{111}^{(1)}q_{1}^3, \\
\bm{\Tilde{F}_{\text{NL}}}^{(1)}(\bm{X}_{2})=\alpha_{11}^{(1)} q_{1}^2-b_{111}^{(1)}q_{1}^3,
\label{eq:forcex}
\end{align}
which can be solved to obtain $\alpha^{(1)}_{11}$ and $b^{(1)}_{111}$, that are namely the quadratic and cubic uncoupled nonlinear terms for the modal coordinate $q_1$. Similarly, by prescribing the system to move on its second eigenmode $\bm{\phi_2}$ we can obtain $\alpha^{(2)}_{22}$ and $b^{(2)}_{222}$. Next, in order to determine the coupled terms, we use the superposition of the eigenmodes as follows:
\begin{eqnarray}
 \bm{X}_{3}&=&+\bm{\phi}_{1} q_1+\bm{\phi_{2}} q_2, \nonumber \\
 \bm{X}_{4}&=&-\bm{\phi}_{1} q_1 -\bm{\phi_{2}} q_2, \nonumber \\
 \bm{X}_{5}&=&+\bm{\phi}_{1} q_1-\bm{\phi}_{2} q_2,
\label{eq:disp2}
\end{eqnarray}
which results in the following set of 3 equations with $a_{12}$, $b_{112}$, $b_{211}$ as unknowns:

\begin{eqnarray}
\bm{\Tilde{F}_{\text{NL}}}^{(1)}(\bm{X}_{3})&=\alpha_{11}^{(1)} q_{1}^2+b_{111}^{(1)}q_{1}^3 + \alpha_{12}^{(1)}q_{1}q_{2}+\alpha_{22}^{(1)}q_{2}^2 \nonumber \\
&\quad +b_{112}^{(1)}q_{1}^2q_{2}+b_{122}^{(1)}q_{1}q_{2}^2+b_{222}^{(1)}q_{2}^3, \nonumber \\
\bm{\Tilde{F}_{\text{NL}}}^{(1)}(\bm{X}_{4})&=\alpha_{11}^{(1)} q_{1}^2-b_{111}^{(1)}q_{1}^3 + \alpha_{12}^{(1)}q_{1}q_{2}+\alpha_{22}^{(1)}q_{2}^2\nonumber \\
&\quad -b_{112}^{(1)}q_{1}^2q_{2}-b_{122}^{(1)}q_{1}q_{2}^2-b_{222}^{(1)}q_{2}^3,\nonumber \\
\bm{\Tilde{F}_{\text{NL}}}^{(1)}(\bm{X}_{5})&=\alpha_{11}^{(1)} q_{1}^2+b_{111}^{(1)}q_{1}^3-\alpha_{12}^{(1)}q_{1}q_{2}+\alpha_{22}^{(1)}q_{2}^2 \nonumber \\
&\quad -b_{112}^{(1)}q_{1}^2q_{2}+b_{122}^{(1)}q_{1}q_{2}^2-b_{222}^{(1)}q_{2}^3.
\label{eq:forcex}
\end{eqnarray}
Since all of the uncoupled parameters ($\alpha^{(r)}_{jk}$ and $b^{(r)}_{jkl}$ where $j=k=l$) are already known thanks to the previous step, the coupling terms between two modes ($\alpha^{(r)}_{jkl}$ and $b^{(r)}_{jkl}$ where $j=k\neq l$, etc.) can be found by solving these three linearly independent equations (Eq.~\eqref{eq:forcex}). In case of coupling between 3 modes of vibration, the third mode's eigenvector shall also be included in the prescribed displacement, i.e., $\bm{X}_c=\bm{\phi}_{j} q_j +\bm{\phi}_{k} q_k +\bm{\phi}_{l} q_l$, and a similar procedure shall be followed to obtain $b^{(r)}_{jkl}$, where $j\neq k\neq l$. We note that this procedure can be generalized and easily applied to a system with $L$ number of modes.

The number of eigenmodes to consider depends on the complexity of the problem and the dynamic range of interest. In our study, and in order to replicate the nonlinear dynamics observed in Fig.~\ref{fig:Fig1}b, we used 11 out-of-plane modes in a frequency range that spans \SIrange[]{20}{90}{\mega\hertz}. For convenience, we compare in Tab.~\ref{tab:nonlinear_coupling} the coupling terms for the first two axisymmetric modes of an ideal (with uniform tension) nanodrum, obtained analytically  (see Supplementary Information I)~\cite{Keskekler2022} with the FE-based ROM approach explained here. We also provide the quadratic and cubic nonlinear terms of the experimentally tested graphene nanodrum for seven modes of vibration in Tab.~4 of Supplementary Information II.  As additional examples, we also provide the nonlinear ROM parameters extracted from this protocol for various other nanomechanical systems such as nanomechanical strings and rectangular membrane resonators in Supplementary Information II.

It is important to mention that the ROM approach sketched here can also account for the influence of in-plane modes of vibration on the nonlinear terms associated with out-of-plane DOFs. This is of great importance for probing nonlinear stiffness terms with accuracy since it has been shown that neglecting the influence of in-plane modes could result in overestimation of the stiffness~\cite{dejan,alipress}. In the analytical setting, effects of in-plane modes could be condensed into the out-of-plane modes by assuming zero in-plane inertia since they have orders of magnitude higher frequencies, thus from the frame of reference of out-of-plane modes, act almost instantaneously. In this way they can be treated statically and their effects can be condensed into the out-of-plane modes, without having to calculate their inertial effects~\cite{MIGNOLET}. Instead of including the in-plane modes, the FEM method described here can automatically include their effect more efficiently by leaving the in-plane displacements free (instead of fixed) while applying the out-of plane membrane displacement $\bm{X}_c$. As such, in-plane effects are automatically condensed into the nonlinear parameters out-of-plane modes. 
\begin{table}[]
\begin{tabular}{l|ll|ll|}
\cline{2-5}
                                 & \multicolumn{2}{c|}{Analytic}     & \multicolumn{2}{c|}{STEP}          \\ \cline{2-5} 
                                & \multicolumn{1}{c|}{Mode 1} & Mode 2 & \multicolumn{1}{c|}{Mode 1} & Mode 2 \\ \hline
\multicolumn{1}{|l|}{$b_{111}$} & \multicolumn{1}{r|}{2.84}      &  \multicolumn{1}{r|}{-0.57}    & \multicolumn{1}{r|}{2.84}      &  \multicolumn{1}{r|}{-0.57}     \\ \hline
\multicolumn{1}{|l|}{$b_{222}$} & \multicolumn{1}{r|}{-3.32}      & \multicolumn{1}{r|}{22.76}     & \multicolumn{1}{r|}{-3.29}      &  \multicolumn{1}{r|}{22.9}    \\ \hline
\multicolumn{1}{|l|}{$b_{112}$} & \multicolumn{1}{r|}{-1.73}      & \multicolumn{1}{r|}{9.25}      & \multicolumn{1}{r|}{-1.71}      & \multicolumn{1}{r|}{9.27}     \\ \hline
\multicolumn{1}{|l|}{$b_{122}$} & \multicolumn{1}{r|}{9.25}      & \multicolumn{1}{r|}{-9.96}      & \multicolumn{1}{r|}{9.27}      &   \multicolumn{1}{r|}{-9.85}    \\ \hline
\end{tabular}
\centering
	\caption{Comparison of the nonlinear coupling coefficients for the first two axisymmetric modes of the graphene nanodrum obtained by an analytical method and the ROM approach. Coefficients are normalized by $c = Eh/r^2$ where $E$ is the Young modulus, $h$ is the thickness and $r$ is the radius of the membrane. We note that the quadratic coupling terms are all zero for a flat symmetric membrane.}
    \label{tab:nonlinear_coupling}
\end{table}
After the construction of $\bm{\eta}$ and the nonlinear ROM, we incorporate the coupled nonlinear differential equations in a numerical continuation package (AUTO)~\cite{doedel2007auto} and obtain the steady-state response for different drive frequencies and drive levels, i.e., the frequency response (step V in Fig.~\ref{fig:fig2}). We utilize the numerical continuation software also to detect bifurcations in the system, which are crucial for understanding the complex nonlinear dynamics of nanoresonators.

We simulate the ROM for different direct and parametric drive levels, $\Tilde{F}^{(r)}(t)=F^{(r)}_{\mathrm{dir}}+F^{(r)}_{\mathrm{par}}$, where $F^{(r)}_{\mathrm{dir}}= J^{(r)}\beta \cos{(\omega_d t)}$ and $F^{(r)}_{\mathrm{par}}=J^{(r)} q_r \gamma \cos{(\omega_d t)}$, with $\gamma$ denoting the parametric drive intensity, $\beta$ the direct drive intensity, and $\bm{J}$ the force mapping vector. We shall note that in order to obtain the modal forces $J^{(r)}$ we use the Duffing shift in the frequency of the high amplitude saddle-node points per drive level. For simplicity,  in our simulations we use the $Q$ factor of the fundamental mode ($Q=180$) for all the modes.
\begin{figure*}
    \centering
    \includegraphics[width=1\linewidth]{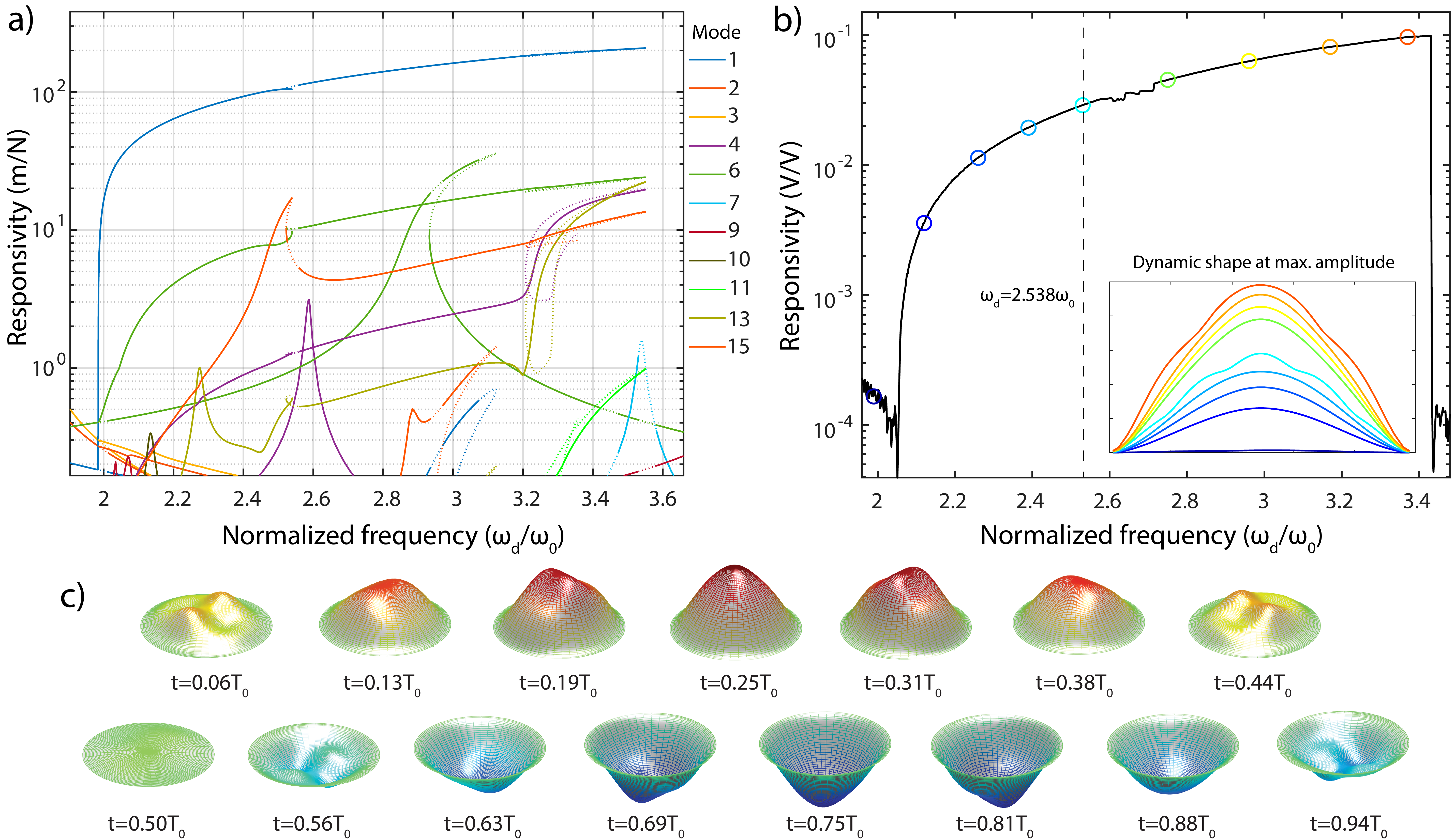}
    \caption{Simulated overall nonlinear dynamic response of the graphene membrane at principal parametric resonance of $f_{0,1}$. a) Simulated shape of the membrane at maximum amplitude level during its parametric resonance at specific drive frequencies, displayed on the experimental parametric resonance curve. Solid lines indicate stable solutions whereas dashed lines indicate unstable ones. b) Full simulated frequency response at the principal parametric resonance of mode $f_{0,1}$, showing activation of many other modes in the system during the strong parametric resonance, specially around the internal resonance region $\omega_d/\omega_0=2.538$. c) Shape of the membrane during a period of oscillation at internal resonance ( $\omega_d/\omega_0=2.538$), displaying multiple mode shapes within a single period of oscillation with different frequencies. $T_0$ is a single period of mode $f_{0,1}$. Amplitude of the response is amplified for visual convenience.}
    \label{fig:shapes}
\end{figure*}

Fig.~\ref{fig:fig3}b shows the simulated frequency response of the nonlinear ROM for various drive levels. Simulations are in good qualitative agreement with the experimental frequency responses in both linear and nonlinear regimes. Although it was attempted to obtain quantitative agreement in the nonlinear regime, this was not fully achieved, possibly due to small imperfections in the membrane that deviate its behavior from the FE model. Similar to experiments, it is possible to observe period doubling bifurcations of modes $f_{0,1}$ and $f_{1,1}$ around $\omega_d=2f_{0,1}$ and $\omega_d=2f_{1,1}$ caused by the parametric drive. When the first parametric resonance reaches the vicinity of the second asymmetric mode $f_{2,1}$, it suffers a reduction in the simulated responsivity in Fig.~\ref{fig:fig3}b which is consistent with the experimental observation in Fig.~\ref{fig:Fig1}b. We also see the frequency locking at the internal resonance. With further increase in the drive level, similar to the experiments, we observe that the frequency locking ``barrier'' is broken, and the frequency of parametric resonance peak surges to 3.5 times the frequency of the fundamental mode $f_{0,1}$. After the surge, we also note the presence of the Neimark-Sacker bifurcation nearby the internal resonance at $\omega_d/\omega_0=2.538$, which indicates the emergence of aperiodic oscillations~\cite{Keskekler2022, Gobat2023}.

The experiments and simulations depicted in Figs.~\ref{fig:Fig1}b and \ref{fig:fig3} demonstrate that, rather than being governed by just two DOFs, the complex motion of the graphene nanodrum around $2.5f_{0,1}$ is a combination of interactions between multiple modes of vibration. We examine the contribution of numerous vibrational modes in the vicinity of the first parametric resonance in order to trace the energy redistribution among various interacting modes. Fig.~\ref{fig:shapes}a shows strong activation of multiple modes at the internal resonance point, where axisymmetric modes $f_{0,2}$ and $f_{0,3}$ (mode numbers 6 and 15) with asymmetric mode $f_{2,1}$ (mode number 4) are most notably excited. Time responses of the modes during one period of $f_{0,1}$ oscillation at $\omega_d/\omega_0 =2.538$ are also shown for convenience (Supplementary Information III). A more visual representation of the time signals can be obtained by using the modal amplitudes $q$ from AUTO to superpose the FEM mode shapes $\bm{\phi}$, thereby reconstructing the total mechanical response and the deflection shape $\bm{X}$ of the graphene nanodrum during internal resonance (see Fig.~\ref{fig:shapes}b and \ref{fig:shapes}c). If we analyze the nanodrum shape at its maximum amplitude level during an oscillation, we can see that the effective deflection shape is unique near the internal resonance point (Fig.~\ref{fig:shapes}b). Dissecting the total motion of the membrane by taking snapshots at different times during a single period of the fundamental mode $f_{0,1}$ (Fig.~\ref{fig:shapes}c) further reveals the strong influence of the multi-modal interaction. It is possible to clearly observe the emergence of other mode shapes during a single oscillation of $f_{0,1}$ parametric resonance. These simulations clearly showcase the energy pathways that lead to the aforementioned nonlinear dissipation phenomenon, not only at the clearly visible internal resonance, but also at responses that look regular, like the direct resonance of $f_{0,1}$. When the global frequency response per mode is analyzed, it is also possible to see the autoparametric activation of multiple modes around the $f_{0,1}$ where most of the energy ends up in $f_{0,3}$ (see Supplementary Information III).

 We note that the favored energy pathways for each system will be distinct due to variations in pre-tension, geometry, and material properties. These physical parameters dictate the system's capacity to ``internally resonate'', due to their effects on the nonlinear terms and resonance frequencies. In the literature, it is common to model nonlinear systems by disregarding the effects of multi-modal interactions, especially if there are no other visible modes contributing to the measurement data. Most of the time, as discussed before, this results in using empirical fit parameters to explain the observations. In this case, the assumption is that, all the interactions effectively renormalize the terms in the single mode equation (generally being Duffing or Duffing-van der Pol). The downfall of this assumption is that, in reality, the effects of multi-modal interactions are amplitude and drive frequency dependent~\cite{Keskekler2021}, whereas re-normalization through empirical fit parameters assumes constant effects. This means that such simplistic models, at best, will agree with the experiments only for a snapshot of frequency response and cannot explain the overall dynamics at higher drive levels and for wide frequency ranges. Utilizing a method that fully relies on physical parameters and includes as many modes as needed, automatically resolves this problem, while clearly displaying the enigmatic nature of these intermodal interactions and energy dissipation pathways.

In summary, we utilized a nonlinear ROM technique to characterize the multi-modal interactions of nanomechanical resonators. We used FEM simulations as the basis to develop our physics-based model, that relies purely on measurable quantities from experiments. We calculated the linear and nonlinear internal forces using FEM simulations to extract quadratic and cubic nonlinear terms for constructing the full nonlinear ROM. By simulating the response curves with the model and comparing the results to nonlinear dynamic measurements of a graphene nanodrum resonator, we showed that the model can replicate complex nonlinear intermodal interactions. Moreover, by tracking simultaneous activation of modal amplitudes, we have identified intermodal energy transfer pathways mediated by nonlinear couplings between multiple modes of vibration. Our study provides an efficient and accurate protocol for modelling complex nonlinear dynamics of nanomechanical resonators in a global manner, purely based on material and geometrical parameters. As a result, we anticipate that this protocol will not only aid in explaining the multi-mode nonlinear dynamics of nanoresonators, but will also serve as a framework for designing optimized nanoresonators that can take advantage of the powerful phenomena that nonlinear dynamics has to offer~\cite{steveopt2015,steveopt2017}.
\bibliographystyle{ieeetr}
\bibliography{multimodal.bib}

\begin{thebibliography}{10}

\bibitem{Chaste2012}
J.~Chaste, A.~Eichler, J.~Moser, G.~Ceballos, R.~Rurali, and A.~Bachtold, ``A
  nanomechanical mass sensor with yoctogram resolution,'' {\em Nature
  Nanotechnology}, vol.~7, pp.~301--304, May 2012.

\bibitem{Fogliano2021}
F.~Fogliano, B.~Besga, A.~Reigue, L.~Mercier~de L{\'e}pinay, P.~Heringlake,
  C.~Gouriou, E.~Eyraud, W.~Wernsdorfer, B.~Pigeau, and O.~Arcizet,
  ``Ultrasensitive nano-optomechanical force sensor operated at dilution
  temperatures,'' {\em Nature Communications}, vol.~12, p.~4124, Jul 2021.

\bibitem{roslon2022probing}
I.~E. Ros{\l}o{\'n}, A.~Japaridze, P.~G. Steeneken, C.~Dekker, and F.~Alijani,
  ``Probing nanomotion of single bacteria with graphene drums,'' {\em Nature
  Nanotechnology}, vol.~17, no.~6, pp.~637--642, 2022.

\bibitem{bachtold2022mesoscopic}
A.~Bachtold, J.~Moser, and M.~Dykman, ``Mesoscopic physics of nanomechanical
  systems,'' {\em Reviews of Modern Physics}, vol.~94, no.~4, p.~045005, 2022.

\bibitem{Steeneken2021}
P.~G. Steeneken, R.~J. Dolleman, D.~Davidovikj, F.~Alijani, and H.~S.~J.
  van~der Zant, ``Dynamics of 2d material membranes,'' {\em 2D Materials},
  vol.~8, p.~042001, Aug 2021.

\bibitem{lifshitzcross}
R.~Lifshitz and M.~C. Cross, {\em Nonlinear Dynamics of Nanomechanical and
  Micromechanical Resonators}, ch.~1, pp.~1--52.
\newblock John Wiley \& Sons, Ltd, 2008.

\bibitem{NayfehOscillations}
A.~H. Nayfeh and D.~T. Mook, {\em Nonlinear Oscillations}, vol.~57 of {\em
  Wiley Classics Library}.
\newblock Wiley-VCH, Mar 1995.

\bibitem{Guttinger2017}
J.~G{\"{u}}ttinger, A.~Noury, P.~Weber, A.~M. Eriksson, C.~Lagoin, J.~Moser,
  C.~Eichler, A.~Wallraff, A.~Isacsson, and A.~Bachtold, ``{Energy-dependent
  path of dissipation in nanomechanical resonators},'' {\em Nature
  Nanotechnology}, vol.~12, pp.~631--636, jul 2017.

\bibitem{Keskekler2021}
A.~Ke{\c{s}}kekler, O.~Shoshani, M.~Lee, H.~S.~J. van~der Zant, P.~G.
  Steeneken, and F.~Alijani, ``Tuning nonlinear damping in graphene
  nanoresonators by parametric--direct internal resonance,'' {\em Nature
  Communications}, vol.~12, p.~1099, Feb 2021.

\bibitem{antonio2012frequency}
D.~Antonio, D.~H. Zanette, and D.~L{\'o}pez, ``Frequency stabilization in
  nonlinear micromechanical oscillators,'' {\em Nature communications}, vol.~3,
  no.~1, pp.~1--6, 2012.

\bibitem{periodtriple}
M.~Wang, D.~J. Perez-Morelo, D.~Lopez, and V.~A. Aksyuk, ``Persistent nonlinear
  phase-locking and nonmonotonic energy dissipation in micromechanical
  resonators,'' {\em Phys. Rev. X}, vol.~12, p.~041025, Dec 2022.

\bibitem{coherentEnergy}
C.~Chen, D.~H. Zanette, D.~A. Czaplewski, S.~Shaw, and D.~L{\'o}pez, ``Direct
  observation of coherent energy transfer in nonlinear micromechanical
  oscillators,'' {\em Nature Communications}, vol.~8, p.~15523, May 2017.

\bibitem{Keskekler2022}
A.~Ke{\c{s}}kekler, H.~Arjmandi-Tash, P.~G. Steeneken, and F.~Alijani,
  ``Symmetry-breaking-induced frequency combs in graphene resonators,'' {\em
  Nano Letters}, vol.~22, pp.~6048--6054, Aug 2022.

\bibitem{oriComb}
D.~A. Czaplewski, C.~Chen, D.~Lopez, O.~Shoshani, A.~M. Eriksson, S.~Strachan,
  and S.~W. Shaw, ``Bifurcation generated mechanical frequency comb,'' {\em
  Phys. Rev. Lett.}, vol.~121, p.~244302, Dec 2018.

\bibitem{markBSpert}
J.~S. Ochs, G.~Rastelli, M.~Seitner, M.~I. Dykman, and E.~M. Weig, ``Resonant
  nonlinear response of a nanomechanical system with broken symmetry,'' {\em
  Phys. Rev. B}, vol.~104, p.~155434, Oct 2021.

\bibitem{dejan}
D.~Davidovikj, F.~Alijani, S.~J. Cartamil-Bueno, H.~S.~J. van~der Zant,
  M.~Amabili, and P.~G. Steeneken, ``Nonlinear dynamic characterization of
  two-dimensional materials,'' {\em Nat Commun}, vol.~8, p.~1253, 2017.

\bibitem{FPUgraph}
D.~Midtvedt, A.~Croy, A.~Isacsson, Z.~Qi, and H.~S. Park, ``Fermi-pasta-ulam
  physics with nanomechanical graphene resonators: Intrinsic relaxation and
  thermalization from flexural mode coupling,'' {\em Phys. Rev. Lett.},
  vol.~112, p.~145503, Apr 2014.

\bibitem{SAJADI}
B.~Sajadi, S.~Wahls, S.~van Hemert, P.~Belardinelli, P.~G. Steeneken, and
  F.~Alijani, ``Nonlinear dynamic identification of graphene's elastic modulus
  via reduced order modeling of atomistic simulations,'' {\em Journal of the
  Mechanics and Physics of Solids}, vol.~122, pp.~161--176, 2019.

\bibitem{dolleman2018opto}
R.~J. Dolleman, S.~Houri, A.~Chandrashekar, F.~Alijani, H.~S. Van Der~Zant, and
  P.~G. Steeneken, ``Opto-thermally excited multimode parametric resonance in
  graphene membranes,'' {\em Scientific reports}, vol.~8, no.~1, pp.~1--7,
  2018.

\bibitem{eichler2011}
A.~Eichler, J.~Moser, J.~Chaste, M.~Zdrojek, I.~Wilson-Rae, and A.~Bachtold,
  ``Nonlinear damping in mechanical resonators made from carbon nanotubes and
  graphene,'' {\em Nature Nanotechnology}, vol.~6, pp.~339--342, Jun 2011.

\bibitem{MURAVYOV20031513}
A.~A. Muravyov and S.~A. Rizzi, ``Determination of nonlinear stiffness with
  application to random vibration of geometrically nonlinear structures,'' {\em
  Computers \& Structures}, vol.~81, no.~15, pp.~1513 -- 1523, 2003.

\bibitem{SAJADIbending}
B.~Sajadi, S.~{van Hemert}, B.~Arash, P.~Belardinelli, P.~G. Steeneken, and
  F.~Alijani, ``Size- and temperature-dependent bending rigidity of graphene
  using modal analysis,'' {\em Carbon}, vol.~139, pp.~334--341, 2018.

\bibitem{gomezmech2015}
A.~Castellanos-Gomez, V.~Singh, H.~S.~J. van~der Zant, and G.~A. Steele,
  ``Mechanics of freely-suspended ultrathin layered materials,'' {\em Annalen
  der Physik}, vol.~527, no.~1-2, pp.~27--44, 2015.

\bibitem{Isacsson_2017}
A.~Isacsson, A.~W. Cummings, L.~Colombo, L.~Colombo, J.~M. Kinaret, and
  S.~Roche, ``Scaling properties of polycrystalline graphene: a review,'' {\em
  2D Materials}, vol.~4, p.~012002, dec 2016.

\bibitem{alipress}
A.~Sarafraz, A.~Givois, I.~Roslon, H.~Liu, H.~Brahmi, G.~Verbiest, P.~G.
  Steeneken, and F.~Alijani, ``Dynamics of pressurized ultra-thin membranes,''
  {\em arXiv.org}, Dec 2022.

\bibitem{MIGNOLET}
M.~P. Mignolet, A.~Przekop, S.~A. Rizzi, and S.~M. Spottswood, ``A review of
  indirect/non-intrusive reduced order modeling of nonlinear geometric
  structures,'' {\em Journal of Sound and Vibration}, vol.~332, no.~10,
  pp.~2437--2460, 2013.

\bibitem{doedel2007auto}
E.~J. Doedel, A.~R. Champneys, F.~Dercole, T.~F. Fairgrieve, Y.~A. Kuznetsov,
  B.~Oldeman, R.~Paffenroth, B.~Sandstede, X.~Wang, and C.~Zhang, ``Auto-07p:
  Continuation and bifurcation software for ordinary differential equations,''
  2007.

\bibitem{Gobat2023}
G.~Gobat, V.~Zega, P.~Fedeli, C.~Touz{\'e}, and A.~Frangi, ``Frequency combs in
  a mems resonator featuring 1:2 internal resonance: ab initio reduced order
  modelling and experimental validation,'' {\em Nonlinear Dynamics}, vol.~111,
  pp.~2991--3017, Feb 2023.

\bibitem{steveopt2015}
S.~Dou, B.~S. Strachan, S.~W. Shaw, and J.~S. Jensen, ``Structural optimization
  for nonlinear dynamic response,'' {\em Philosophical Transactions of the
  Royal Society A: Mathematical, Physical and Engineering Sciences}, vol.~373,
  no.~2051, p.~20140408, 2015.

\bibitem{steveopt2017}
L.~L. Li, P.~M. Polunin, S.~Dou, O.~Shoshani, B.~Scott~Strachan, J.~S. Jensen,
  S.~W. Shaw, and K.~L. Turner, ``Tailoring the nonlinear response of mems
  resonators using shape optimization,'' {\em Applied Physics Letters},
  vol.~110, no.~8, p.~081902, 2017.

\end{thebibliography}
\onecolumngrid
\clearpage


\section*{Supplementary material}

# Supplementary Information

# Characterizing multi-mode nonlinear dynamics of nanomechanical resonators

Ata Keşkekler[*,1], Vincent Bos[1], Alejandro M. Aragón[1], Peter G. Steeneken[1,2], and Farbod Alijani[*,1]

[1]Department of Precision and Microsystems Engineering, TU Delft, The Netherlands
[2]Kavli Institute of Nanoscience, TU Delft, The Netherlands

April 3, 2023

# Contents

[*]Corresponding authors:
Ata Keşkekler <a.keskekler-1@tudelft.nl>,
Farbod Alijani <f.alijani@tudelft.nl>

# 1 Analytical model for ideal membrane with two-degrees-of-freedom

In order to benchmark the method provided in the main text, we use the Rayleigh-Ritz approach to obtain the equations of motion for a circular membrane [1, 2, 3]. The elastic strain energy of the drum is obtained as

$$U = \int_0^{2\pi} \int_0^R \frac{Eh}{2(1-\nu^2)} \left( \epsilon_{rr}^2 + \epsilon_{\theta\theta}^2 + 2\nu\epsilon_{rr}\epsilon_{\theta\theta} + \frac{1-\nu}{2}\gamma_{r\theta}^2 \right) r dr d\theta, \quad (1)$$

in which $h$ is the thickness of the drum, $R$ is the radius, $E$ is the Young's modulus, and $\nu$ is the Poisson's ratio. Moreover, $\epsilon_{rr}$, $\epsilon_{\theta\theta}$, and $\gamma_{r\theta}$ are the normal and shear strains. During axisymmetric oscillations $\gamma_{r\theta} = 0$, and the normal strains are obtained in terms of the transverse deflection ($w$) and radial displacement ($u$) of the drum as follows

$$\epsilon_{rr} = \frac{\partial u}{\partial r} + \frac{1}{2}\left(\frac{\partial w}{\partial r}\right)^2 + \left(\frac{\partial w}{\partial r}\right)\left(\frac{\partial w_0}{\partial r}\right), \quad (2)$$

$$\epsilon_{\theta\theta} = \frac{u}{r}. \quad (3)$$

In Eq.(2), $w_0$ is the initial transverse displacement of the drum associated with zero initial stress, and is equal to zero when assuming ideal symmetric membrane. By taking into account the modal interactions between the first two axisymmetric eigenmodes and fixed boundary conditions ($u = w = 0$), the solution is approximated as

$$w = q_1(t) J_0\left(Z_1 \frac{r}{R}\right) + q_2(t) J_0\left(Z_2 \frac{r}{R}\right), \quad (4)$$

$$u = u_0 r + r(R-r) \sum_{j=1}^n y_j(t) r^{j-1}, \quad (5)$$

where $u_0$ is the initial radial displacement due to pre-tension $n_0$, and $q_1$ and $q_2$ are generalized coordinates associated with the first and the second axisymmetric mode of the drum, respectively. Moreover, $J_0$ is the Bessel function of order zero and $Z_1$ and $Z_2$ are its first two roots. We Taylor approximate $J_0\left(Z_1 \frac{r}{R}\right)$ to 10th order and $J_0\left(Z_2 \frac{r}{R}\right)$ to 14th order to accurately capture the mode shapes. In addition, $y_j$ are the radial generalized coordinates and $n$ is the number of these coordinates retained in the approximation where convergence is achieved at $n = 8$.

The kinetic energy of the drum is then obtained as

$$T = \frac{1}{2}\rho h \int_0^{2\pi} \int_0^R \dot{w}^2 r dr d\theta, \quad (6)$$

in which the overdot represents differentiation with respect to time $t$, and $\rho$ is the mass density.

Next, we obtain Lagrange equations as follows:

$$\frac{\partial \mathcal{L}}{\partial q_i} - \frac{d}{dt}\left(\frac{\partial \mathcal{L}}{\partial \dot{q}_i}\right) = 0, \quad (7)$$

where $\mathcal{L}$ is the Lagrangian ($\mathcal{L} = T - U$), resulting in the equations:

$$m_1 \ddot{q}_1 + k_1 q_1 + \alpha_{11}^{(1)} q_1^2 + \alpha_{12}^{(1)} q_1 q_2 + \alpha_{22}^{(1)} q_2^2 + b_{111}^{(1)} q_1^3 + b_{112}^{(1)} q_1^2 q_2 + b_{122}^{(1)} q_1 q_2^2 + b_{222}^{(1)} q_2^3 = 0, \quad (8)$$

$$m_2 \ddot{q}_2 + k_2 q_2 + \alpha_{11}^{(2)} q_1^2 + \alpha_{12}^{(2)} q_1 q_2 + \alpha_{22}^{(2)} q_2^2 + b_{111}^{(2)} q_1^3 + b_{112}^{(2)} q_1^2 q_2 + b_{122}^{(2)} q_1 q_2^2 + b_{222}^{(2)} q_2^3 = 0, \quad (9)$$

in which $\alpha_{jk}^{(r)}$ and $b_{jkl}^{(r)}$ are the quadratic and cubic nonlinear terms. It is possible to express the coefficients in the equations of motion in terms of mechanical and geometric properties of the drum as follows

$$\omega_1 = \sqrt{\frac{k_1}{m_1}} = \frac{2.4}{R}\sqrt{\frac{n_0}{\rho h}} \tag{10a}$$

$$\omega_2 = \sqrt{\frac{k_2}{m_2}} = \frac{5.5}{R}\sqrt{\frac{n_0}{\rho h}}, \tag{10b}$$

$$b_{111}^{(1)} = 2.84\frac{Eh}{R^2}, b_{111}^{(2)} = -0.57\frac{Eh}{R^2} \tag{10c}$$

$$b_{222}^{(1)} = -3.32\frac{Eh}{R^2}, b_{222}^{(2)} = 22.76\frac{Eh}{R^2} \tag{10d}$$

$$b_{112}^{(1)} = -1.73\frac{Eh}{R^2}, b_{112}^{(2)} = 9.25\frac{Eh}{R^2} \tag{10e}$$

$$b_{122}^{(1)} = 9.25\frac{Eh}{R^2}, b_{222}^{(2)} = -9.96\frac{Eh}{R^2} \tag{10f}$$

in which the coefficients were evaluated assuming $\nu = 0.16$. The quadratic terms $\alpha_{jk}^{(r)}$ are all zero due to the ideal symmetric membrane assumption ($w_0 = 0$).

# 2 Nonlinear reduced-order model parameters for different nanomechanical resonators

Here we provide the nonlinear reduced-order model parameters extracted using the STEP method [4], for the graphene nanodrum discussed in the main text, as well as a rectangular membrane and a string resonator. In the following tables, it is possible to find the definitions of the physical parameters for each model, dimensional prefactors for the terms in the nonlinear reduced-order model, and the dimensionless coefficients for the nonlinear terms. We note that for flat ideal circular, square and string resonators, the quadratic nonlinear terms are nonexistent for out-of-plane modes in the absence of in-plane modes.

## 2.1 Reduced-order model parameters for the experimental graphene drum

Reduced-order model of the graphene drum is based on a finite element (FE) model constructed in COMSOL, considering fixed boundary condition with plate elements. Eigenmodes of the graphene drum can be found in Supplementary Fig. 1. The description of model parameters, scaling values and linear modal parameters can be found in Supplementary Tables 1, 2 and 3. We construct the model using the first three axisymmetric and first four asymmetric out-of-plane modes of the structure, excluding their degenerates. Nonlinear coefficients can be found in Supplementary Table 4.

Supplementary Table 1: Parameters for the graphene drum resonator

| Parameter | Formula | Units | Description |
|---|---|---|---|
| $E$ | - | Pa | Young's modulus |
| $h$ | - | m | Thickness |
| $R$ | - | m | Radius |
| $\nu$ | - | - | Poisson's ratio |
| $\rho$ | - | kg/m$^3$ | Density |
| $\sigma$ | - | Pa | Stress |
| $\tilde{\rho}$ | $\rho h$ | kg/m$^2$ | Mass per unit area |
| $T_0$ | $\sigma h$ | N m$^{-1}$ | Membrane tension |

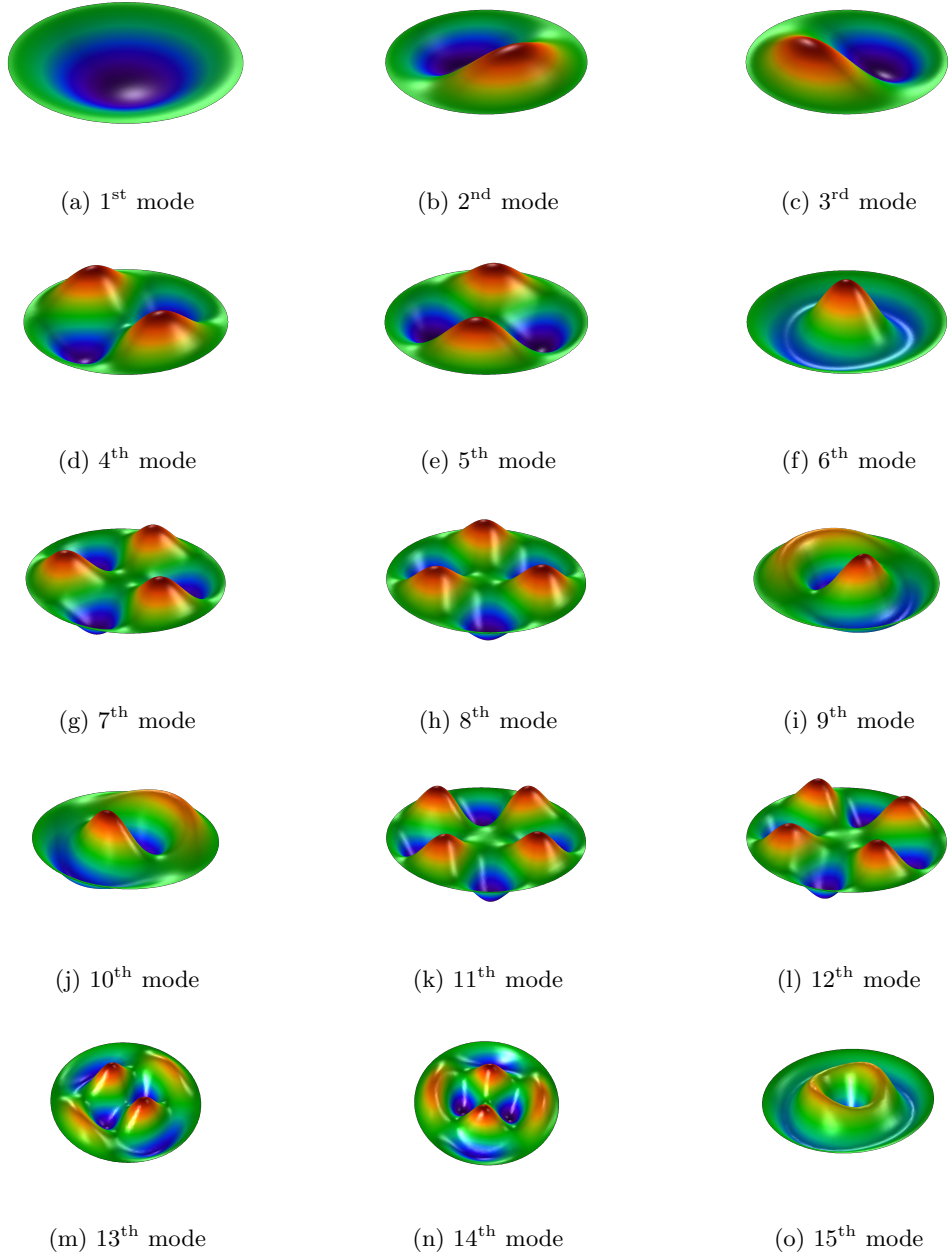

Supplementary Figure 1: Eigenmodes of the graphene drum resonator

Supplementary Table 2: Scaling values for the graphene drum equations

|  | Linear | | | Nonlinear | |
|---|---|---|---|---|---|
| $m_\mathrm{t}$ | $k_0$ | $\omega_0$ | | $a_\mathrm{dim}$ | $b_\mathrm{dim}$ |
| $\pi R^2 h \rho$ | $T_0$ | $\dfrac{2.405}{R}\sqrt{\dfrac{T_0}{\tilde{\rho}}}$ | | $[-]$ | $\dfrac{Eh}{R^2}$ |

Supplementary Table 3: Dimensionless linear modal values for the experimental graphene drum from the main text. Eigenfrequencies are normalized w.r.t. the fundamental mode ($\omega_1$). Modes: 1,2,4,6,7,9,15

|  | Eq. 1 | Eq. 2 | Eq. 3 | Eq. 4 | Eq. 5 | Eq. 6 | Eq. 7 |
|---|---|---|---|---|---|---|---|
| $m/m_t$ | 0.204 | 0.195 | 0.183 | 0.103 | 0.192 | 0.118 | 0.0698 |
| $k/k_0$ | 5.78 | 16.5 | 34.6 | 24.2 | 68.3 | 55.4 | 64.9 |
| $\omega/\omega_1$ | 1 | 1.73 | 2.59 | 2.89 | 3.54 | 4.07 | 5.73 |

Supplementary Table 4: Dimensionless nonlinear coefficients for the graphene drum ($\nu = 0.16$) from the main text. Modes: 1,2,4,6,7,9,15.

| | Eq. 1 | Eq. 2 | Eq. 3 | Eq. 4 | Eq. 5 | Eq. 6 | Eq. 7 |
|---|---|---|---|---|---|---|---|
| $b_{111}$ | 2.55 | 0 | 0.313 | 1.34 | 0.000204 | 0.000435 | −0.529 |
| $b_{112}$ | 0 | 8.57 | −0.000307 | −0.000596 | −0.426 | −4.05 | 0.000196 |
| $b_{113}$ | 0.942 | −0.000283 | 9.79 | 1.07 | −0.000111 | 0.000244 | −0.0467 |
| $b_{114}$ | 4.02 | −0.000585 | 1.07 | 10.8 | 0.00124 | −0.000745 | 2.96 |
| $b_{115}$ | 0.00058 | −0.434 | −0.000109 | 0.00124 | 12 | 0.71 | 0.000527 |
| $b_{116}$ | 0.00128 | −4.05 | 0.000289 | −0.000762 | 0.711 | 15.9 | 0.0028 |
| $b_{117}$ | −1.61 | 0.000168 | −0.0469 | 2.96 | 0.000479 | 0.0028 | 14.2 |
| $b_{122}$ | 8.61 | 0 | −5.82 | 3.82 | 0.00154 | 0.00208 | −2.56 |
| $b_{123}$ | −0.143 | −11.8 | 0.349 | 0.0771 | −13.8 | 7.28 | 0.0707 |
| $b_{124}$ | 0.00129 | 7.69 | 0.0838 | 0.167 | 0.621 | −14.5 | 0.08 |
| $b_{125}$ | −0.826 | 0.174 | −13.7 | 0.625 | 0.434 | 0.0744 | 2.17 |
| $b_{126}$ | −8.12 | 0.0522 | 7.31 | −14.5 | 0.0669 | 0.26 | 0.184 |
| $b_{127}$ | −0.0667 | −5.07 | 0.0861 | 0.0629 | 2.17 | 0.158 | 0.361 |
| $b_{133}$ | 9.88 | 0.000849 | 1.46 | 1.37 | −0.000543 | 0.00217 | −7.65 |
| $b_{134}$ | 2.13 | 0.000409 | 2.83 | 6.92 | 0.00248 | −0.000509 | 4.82 |
| $b_{135}$ | −0.0277 | −13.7 | 0.22 | −0.000755 | 1.95 | 6.94 | −0.0145 |
| $b_{136}$ | −0.00152 | 7.33 | 0.085 | 0.019 | 6.94 | −9.79 | −0.00623 |
| $b_{137}$ | −0.0863 | −0.000881 | −15.3 | 4.82 | 0.0112 | −0.00304 | 10.5 |
| $b_{144}$ | 10.8 | −0.000906 | 3.54 | 17.4 | 0.00168 | −0.000612 | 8.08 |
| $b_{145}$ | 0.0254 | 0.634 | 0.0135 | 0.133 | −1.84 | 2.25 | −0.0204 |
| $b_{146}$ | 0.0829 | −14.4 | 0.0383 | 0.167 | 2.25 | 34 | −0.00671 |
| $b_{147}$ | 5.92 | −0.00501 | 4.86 | 16.3 | 0.00322 | 0.0132 | 28.8 |
| $b_{155}$ | 12.1 | −0.000487 | 1.18 | −1.01 | −0.00161 | 0.00579 | −11.1 |
| $b_{156}$ | 1.45 | 0.02 | 7.02 | 2.27 | 0.197 | 0.293 | −0.336 |
| $b_{157}$ | −0.032 | 2.16 | −0.000708 | −0.0264 | −22.1 | −0.336 | 0.136 |
| $b_{166}$ | 16.1 | 0.00254 | −4.79 | 16.8 | 0.00614 | −0.0112 | 2.66 |
| $b_{167}$ | −0.033 | 0.161 | −0.00144 | −0.0053 | −0.335 | 5.38 | 0.144 |
| $b_{177}$ | 14.4 | −0.00684 | 5.31 | 14.3 | 0.00216 | 0.0112 | 3.13 |
| $b_{222}$ | 0 | 18.7 | −0.000908 | −0.00143 | 1.94 | −6.99 | 0.000362 |
| $b_{223}$ | −5.73 | −0.00272 | 45.2 | 6.43 | −0.00445 | −0.00177 | 6.38 |
| $b_{224}$ | 3.83 | −0.00433 | 6.42 | 16 | 0.00159 | −0.00127 | 3.57 |
| $b_{225}$ | 0.00146 | 5.81 | −0.00434 | 0.00156 | 47.7 | 4.2 | −0.00925 |
| $b_{226}$ | 0.00194 | −21 | −0.00157 | −0.00127 | 4.21 | 53 | 0.0126 |
| $b_{227}$ | −2.48 | 0.000901 | 6.38 | 3.55 | −0.00939 | 0.0125 | 22.2 |
| $b_{233}$ | 0.00075 | 45.2 | −0.00485 | −0.00484 | 31.9 | −5.51 | −0.00498 |
| $b_{234}$ | 0.00452 | 12.9 | −0.00523 | −0.0825 | 20.3 | 7.73 | −0.0346 |
| $b_{235}$ | −13.5 | −0.224 | 64 | 20.3 | −0.399 | −0.0729 | 15.6 |
| $b_{236}$ | 7.3 | 0.0157 | −10.9 | 7.74 | −0.0876 | −0.285 | −9.19 |
| $b_{237}$ | −0.0112 | 12.8 | −0.0207 | −0.0566 | 15.6 | −9.18 | −0.158 |
| $b_{244}$ | −0.000933 | 16 | −0.00287 | −0.00374 | −3.94 | −14.1 | −0.00715 |
| $b_{245}$ | 0.605 | 0.00898 | 20.3 | −7.81 | 0.0907 | 0.00222 | −2.75 |
| $b_{246}$ | −14.3 | −0.0382 | 7.73 | −28.1 | 0.00247 | 0.15 | −11.7 |
| $b_{247}$ | 0.0374 | 7.29 | 0.0448 | 0.237 | −2.76 | −11.8 | 0.376 |
| $b_{255}$ | −0.000489 | 47.8 | 0.00469 | −0.00145 | 0.0154 | 10.7 | −0.012 |
| $b_{256}$ | −0.00483 | 8.51 | −0.00451 | −0.0111 | 21.5 | 5.05 | 0.000509 |
| $b_{257}$ | 2.13 | −0.0207 | 15.6 | −2.74 | 0.00764 | 0.00618 | −12.7 |
| $b_{266}$ | 0.00243 | 53.3 | −0.0148 | 0.00474 | 2.42 | −66.2 | −0.018 |
| $b_{267}$ | −0.108 | −0.0164 | −9.17 | −11.7 | 0.0066 | 0.0415 | −29.3 |
| $b_{277}$ | −0.00664 | 22.3 | −0.00488 | −0.0109 | −6.38 | −14.7 | −0.00534 |
| $b_{333}$ | 0.476 | −0.00144 | 44.9 | 4.06 | −0.00568 | −0.00114 | 0.176 |
| $b_{334}$ | 1.35 | −0.00473 | 12.2 | 45.4 | −0.000223 | −0.0106 | 8.8 |
| $b_{335}$ | −0.000782 | 31.8 | −0.0168 | −0.000288 | 121 | 11.2 | −0.00672 |
| $b_{336}$ | 0.00211 | −5.46 | −0.00299 | −0.0106 | 11.2 | 54.8 | −0.00819 |
| $b_{337}$ | −7.59 | −0.00504 | 0.53 | 8.78 | −0.00705 | −0.00823 | 45.1 |
| $b_{344}$ | 3.5 | −0.00279 | 45.4 | 0.37 | −0.00526 | −0.00348 | 5.32 |
| $b_{345}$ | −0.00534 | 20.2 | 0.0276 | −0.0768 | 5.01 | 58.3 | −0.0532 |
| $b_{346}$ | −0.00706 | 7.68 | −0.00618 | −0.0376 | 58.3 | 12.7 | 0.00249 |
| $b_{347}$ | 4.81 | −0.000293 | 17.6 | 10.5 | −0.0412 | 0.0192 | 4.01 |
| $b_{355}$ | 1.15 | 0.00491 | 121 | 2.55 | 0.000372 | −0.002 | −6.49 |
| $b_{356}$ | 6.9 | −0.0452 | 22.4 | 58.3 | −0.145 | −0.213 | −14 |
| $b_{357}$ | −0.00073 | 15.5 | 0.00302 | −0.0546 | −13 | −14 | −0.085 |
| $b_{366}$ | −4.85 | −0.0145 | 54.9 | 6.4 | −0.0298 | 0.00973 | 11.9 |
| $b_{367}$ | −0.00578 | −9.12 | −0.00329 | 0.00184 | −14 | 23.8 | −0.0382 |
| $b_{377}$ | 5.24 | −0.00472 | 45.1 | 2.09 | −0.00347 | 0.0116 | −2.47 |
| $b_{444}$ | 5.72 | −0.00118 | 0.121 | 26.5 | 0.00194 | −0.00152 | 13.7 |
| $b_{445}$ | 0.00157 | −3.92 | −0.00516 | 0.00524 | 57.3 | −4.5 | 0.0103 |
| $b_{446}$ | −0.000414 | −14 | −0.00338 | −0.00593 | −4.5 | 61.5 | 0.0181 |
| $b_{447}$ | 8.05 | −0.00703 | 5.31 | 41.1 | 0.0101 | 0.0182 | 70.8 |
| $b_{455}$ | −1.05 | −0.00139 | 2.54 | 57.3 | 0.00456 | −0.00335 | −11.8 |
| $b_{456}$ | 2.22 | −0.00729 | 58.3 | −8.97 | 0.0408 | 0.0648 | 8.01 |
| $b_{457}$ | 0.0154 | −2.73 | −0.0417 | 0.0565 | −23.6 | 8.01 | 0.0482 |
| $b_{466}$ | 16.5 | 0.00502 | 6.39 | 61.5 | −0.00738 | −0.0574 | 24.7 |
| $b_{467}$ | 0.0212 | −11.6 | 0.0228 | 0.0718 | 8.01 | 49.5 | 0.129 |
| $b_{477}$ | 14.1 | −0.0106 | 2.08 | 70.8 | −0.00654 | 0.0288 | 77.4 |
| $b_{555}$ | −0.000543 | 0.0125 | 0.000385 | 0.00193 | 116 | 2.28 | −0.00666 |
| $b_{556}$ | 0.00566 | 10.7 | −0.00135 | −0.00352 | 6.83 | 105 | −0.0293 |
| $b_{557}$ | −10.9 | −0.012 | −6.49 | −11.8 | −0.0234 | −0.0292 | 61.7 |
| $b_{566}$ | 0.00617 | 2.45 | −0.0296 | −0.00728 | 105 | 42.2 | −0.017 |
| $b_{567}$ | −0.308 | −0.00115 | −14 | 8.01 | −0.0505 | −0.0185 | −4.96 |
| $b_{577}$ | 0.00192 | −6.34 | −0.00334 | −0.00672 | 61.8 | −2.49 | 0.0524 |
| $b_{666}$ | −0.00265 | −21.9 | 0.0039 | −0.0176 | 14.1 | 99.4 | 0.0251 |
| $b_{667}$ | 2.89 | −0.0179 | 11.9 | 24.7 | −0.0172 | 0.077 | 67.3 |
| $b_{677}$ | 0.0108 | −14.5 | 0.0118 | 0.0285 | −2.49 | 67.3 | 0.0274 |
| $b_{777}$ | 1.13 | −0.000123 | −0.814 | 25.8 | 0.0211 | 0.00682 | 81.4 |

## 2.2 Reduced-order model parameters for a square membrane resonator

Reduced-order model of an ideal square membrane is based on an FE model constructed in COMSOL, considering pinned boundary conditions with membrane elements. Eigenmodes of the square drum can be found in Supplementary Fig. 2. The description of model parameters, scaling values and linear modal parameters can be found in Supplementary Tables 5, 6 and 7. We construct the model using the first eleven out-of-plane modes of the structure, excluding their degenerates, resulting in seven modes. Nonlinear coefficients can be found in Supplementray Table 8.

Supplementary Table 5: Parameters for a square membrane and string resonators

| Parameter | Formula | Units | Description |
|---|---|---|---|
| $E$ | - | Pa | Young's modulus |
| $h$ | - | m | Thickness |
| $L$ | - | m | Length |
| $\nu$ | - | - | Poisson's ratio |
| $\rho$ | - | kg/m$^3$ | Density |
| $\sigma$ | - | Pa | Stress |
| $\tilde{\rho}$ | $\rho h$ | kg/m$^2$ | Mass per unit area |
| $T_0$ | $\sigma h$ | N m$^{-1}$ | Membrane tension |

Supplementary Table 6: Scaling values for a square membrane resonator

| Linear | | | Nonlinear | |
|---|---|---|---|---|
| $m_t$ | $k_0$ | $\omega_0$ | $a_{\dim}$ | $b_{\dim}$ |
| $L^2 h \rho$ | $T_0$ | $\dfrac{\pi}{L}\sqrt{\dfrac{2T_0}{\tilde{\rho}}}$ | $[-]$ | $\dfrac{\pi}{1.27 - 0.97\nu}\dfrac{Eh}{L^2}$ |

Supplementary Table 7: Dimensionless linear modal values of a square membrane resonator for modes: 1,2,4,5,7,9,11.

| | Eq. 1 | Eq. 2 | Eq. 3 | Eq. 4 | Eq. 5 | Eq. 6 | Eq. 7 |
|---|---|---|---|---|---|---|---|
| $m/m_t$ | 0.25 | 0.215 | 0.25 | 0.206 | 0.154 | 0.227 | 0.248 |
| $k/k_0$ | 4.93 | 10.6 | 19.7 | 20.3 | 19.8 | 38.2 | 44.2 |
| $\omega/\omega_0$ | 1 | 1.58 | 2 | 2.24 | 2.55 | 2.92 | 3 |

Supplementary Table 8: Dimensionless nonlinear coefficients for a square membrane resonator modelled with modes: 1,2,4,5,7,9,11.

| | Eq. 1 | Eq. 2 | Eq. 3 | Eq. 4 | Eq. 5 | Eq. 6 | Eq. 7 |
|---|---|---|---|---|---|---|---|
| $b_{111}$ | 3.65 | 0 | 0 | −0.0807 | 0 | −0.000 131 | −0.0466 |
| $b_{112}$ | 0 | 11.6 | 0 | −0.000 267 | 1.17 | −1.17 | −0.000 234 |
| $b_{113}$ | 0 | 0 | 15.2 | −0.000 269 | 0 | 0 | −0.001 12 |
| $b_{114}$ | −0.242 | 0 | −0.000 418 | 16.4 | −0.000 231 | −0.000 154 | −0.635 |
| $b_{115}$ | 0 | 1.15 | 0.000 103 | −0.000 291 | 12.9 | −1.27 | 0.001 07 |
| $b_{116}$ | 0 | −1.13 | −0.000 234 | −0.000 123 | −1.27 | 31.8 | 0.002 11 |
| $b_{117}$ | −0.144 | −0.000 263 | −0.001 17 | −0.634 | 0.000 812 | 0.001 59 | 27.4 |
| $b_{122}$ | 11.6 | −0.000 387 | 10.4 | 2.87 | −0.000 952 | −0.0015 | 0.367 |
| $b_{123}$ | −0.0597 | 20.7 | −0.303 | −0.0311 | −13.7 | −8.35 | −0.109 |
| $b_{124}$ | −0.0514 | 5.63 | −0.0362 | −0.385 | 4.47 | 7.54 | 0.179 |
| $b_{125}$ | 2.29 | −0.057 | −14.1 | 4.49 | −0.193 | −0.0697 | −18 |
| $b_{126}$ | −2.32 | −0.126 | −8.16 | 8 | −0.0441 | −0.494 | −9.19 |
| $b_{127}$ | 0.0561 | 0.791 | 0.0117 | 0.0806 | −18.2 | −8.97 | −0.397 |
| $b_{133}$ | 15.2 | −0.000 448 | 0.002 01 | 1.2 | 0.002 78 | −0.003 27 | 8.13 |
| $b_{134}$ | −0.019 | 0.001 78 | 2.29 | −0.186 | 0.000 313 | 0.006 95 | −0.0058 |
| $b_{135}$ | −0.0147 | −14 | −0.121 | −0.005 15 | 10.3 | −9.72 | −0.16 |
| $b_{136}$ | −0.0233 | −8.18 | −0.14 | −0.0121 | −9.85 | −3.8 | −0.0321 |
| $b_{137}$ | 0.0116 | 0.001 07 | 16.5 | −0.007 81 | 0.007 65 | −0.006 99 | −0.232 |
| $b_{144}$ | 16.4 | 0 | 0 | 2.26 | −0.001 41 | 0.004 65 | −19.8 |
| $b_{145}$ | −0.0133 | 4.48 | 0.003 81 | −0.175 | 3.1 | −8.76 | 0.0714 |
| $b_{146}$ | −0.0633 | 8.01 | 0.005 51 | −0.26 | −8.47 | 19.2 | 0.227 |
| $b_{147}$ | −1.16 | 0.002 15 | −0.005 16 | −39.3 | −0.000 571 | 0.0115 | −0.279 |
| $b_{155}$ | 12.9 | −0.003 23 | 5.31 | 1.64 | −0.001 01 | −0.001 47 | 2.64 |
| $b_{156}$ | −2.53 | −0.0488 | −9.79 | −8.34 | −0.148 | −0.267 | −17.9 |
| $b_{157}$ | 0.0263 | −18.2 | −0.0387 | 0.0259 | 5.39 | −17.6 | −0.312 |
| $b_{166}$ | 31.8 | 0.0107 | −1.71 | 9.94 | 0.0026 | −0.0312 | 0.307 |
| $b_{167}$ | 0.061 | −9.15 | 0.000 304 | 0.189 | −17.8 | 1 | −0.458 |
| $b_{177}$ | 27.4 | 0 | 0.0117 | 0.2 | 0.0051 | −0.0057 | 0.11 |
| $b_{222}$ | −0.000 165 | 21.9 | 0 | 0.000 103 | −0.117 | −1.95 | 0.000 453 |
| $b_{223}$ | 10.5 | 0.000 308 | 58.2 | 2.24 | 0.002 78 | 0.0003 | 8.15 |
| $b_{224}$ | 2.91 | 0 | 2.24 | 58.7 | −0.003 07 | 0.005 54 | −6.02 |
| $b_{225}$ | −0.000 944 | −0.34 | 0.002 68 | −0.002 86 | 51 | −6.71 | 0.007 85 |
| $b_{226}$ | −0.001 46 | −5.86 | −0.000 207 | 0.004 51 | −6.71 | 79.3 | −0.000 829 |
| $b_{227}$ | 0.307 | 0.001 35 | 8.17 | −6.03 | 0.007 68 | −0.001 78 | 69.9 |
| $b_{233}$ | −0.000 749 | 58.2 | −0.0025 | −0.001 13 | −28.8 | −2.03 | −0.0032 |
| $b_{234}$ | −0.0416 | 4.4 | −0.115 | −0.19 | −7.88 | 37.6 | 0.0978 |
| $b_{235}$ | −14.1 | 0.0428 | −57.7 | −7.9 | −0.224 | 0.0697 | −64.9 |
| $b_{236}$ | −8.21 | −0.0898 | −4.19 | 37.7 | 0.144 | −0.495 | 11.6 |
| $b_{237}$ | −0.008 33 | 16.4 | −0.0144 | 0.0216 | −64.9 | 11.6 | −0.316 |
| $b_{244}$ | 0 | 58.7 | 0.003 45 | 0.002 42 | 34.5 | 21.1 | 0.001 02 |
| $b_{245}$ | 4.56 | −0.0935 | −7.91 | 68.8 | −0.267 | −0.195 | −26.9 |
| $b_{246}$ | 8.12 | −0.169 | 37.8 | 42 | −0.116 | −0.904 | −58.8 |
| $b_{247}$ | 0.0902 | −12 | 0.0513 | 0.398 | −26.9 | −58.7 | −0.63 |
| $b_{255}$ | −0.003 67 | 51 | −0.008 05 | −0.006 74 | −20.2 | 10.4 | −0.0024 |
| $b_{256}$ | −0.000 602 | −13.5 | 0.0222 | −0.0165 | 20.7 | 46 | 0.0198 |
| $b_{257}$ | −18.3 | 0.0616 | −65 | −26.9 | 0.146 | 0.128 | −15.4 |
| $b_{266}$ | 0.0114 | 79.4 | 0.009 16 | 0.0192 | 23.2 | 4.87 | −0.0265 |
| $b_{267}$ | −9.2 | 0.121 | 11.6 | −59.1 | 0.0295 | 0.535 | 72.6 |
| $b_{277}$ | 0.001 43 | 70 | −0.006 57 | −0.02 | −7.5 | 36.4 | −0.0217 |
| $b_{333}$ | 0.000 617 | −0.0005 | 58.5 | −0.002 02 | 0.003 16 | 0.000 402 | 0.004 16 |
| $b_{334}$ | 1.21 | 0 | −0.006 48 | 67 | −0.003 99 | 0.003 73 | 4.26 |
| $b_{335}$ | 0.001 61 | −28.9 | 0.009 64 | −0.004 11 | 96.5 | 3.74 | 0.011 |
| $b_{336}$ | −0.0036 | −2.01 | −0.000 215 | 0.002 01 | 3.72 | 145 | −0.004 39 |
| $b_{337}$ | 8.21 | −0.002 74 | 0.0113 | 4.26 | 0.0106 | −0.004 83 | 156 |
| $b_{344}$ | 0 | 0.002 94 | 67 | 0.001 08 | 0.005 33 | −0.000 892 | −0.007 72 |
| $b_{345}$ | 0.009 29 | −7.9 | −0.0623 | −0.089 | 10.9 | −16.8 | −0.0892 |
| $b_{346}$ | −0.0278 | 37.8 | −0.0846 | −0.116 | −16.8 | 2 | 0.0116 |
| $b_{347}$ | 0.006 47 | 0.0126 | 8.55 | 0.122 | 0.001 04 | −0.0017 | −0.208 |
| $b_{355}$ | 5.3 | −0.008 47 | 96.5 | 5.51 | 0.028 | −0.0255 | 71.2 |
| $b_{356}$ | −9.82 | 0.0311 | 7.38 | −16.9 | −0.134 | −0.235 | −19.5 |
| $b_{357}$ | −0.0299 | −65.1 | 0.005 07 | 0.006 18 | 143 | −19.5 | −0.193 |
| $b_{366}$ | −1.73 | 0.0104 | 145 | 1.16 | −0.0114 | −0.097 | −12.7 |
| $b_{367}$ | −0.0406 | 11.6 | 0.0688 | −0.0194 | −19.6 | −25.1 | −0.472 |
| $b_{377}$ | 0.0107 | −0.007 93 | 156 | 0.0176 | 0.008 23 | −0.0476 | 0.137 |
| $b_{444}$ | 0.767 | 0.000 402 | 0.000 369 | 81.4 | −0.003 72 | 0.0116 | −7.64 |
| $b_{445}$ | −0.001 48 | 34.5 | 0.005 03 | −0.0106 | 112 | 41 | 0.0106 |
| $b_{446}$ | 0.002 54 | 21.2 | −0.003 15 | 0.0327 | 40.9 | 192 | −0.0431 |
| $b_{447}$ | −20.1 | 0.003 05 | −0.007 74 | −22.9 | 0.0118 | −0.0465 | 253 |
| $b_{455}$ | 1.66 | −0.008 14 | 5.52 | 112 | −0.008 63 | −0.0149 | −9.73 |
| $b_{456}$ | −8.43 | −0.114 | −16.9 | 81.7 | −0.224 | −0.459 | −80.8 |
| $b_{457}$ | 0.0719 | −26.9 | −0.004 88 | 0.178 | −19.3 | −80.8 | −0.367 |
| $b_{466}$ | 9.98 | 0.0206 | 1.16 | 192 | 0.0156 | 0.393 | −103 |
| $b_{467}$ | 0.195 | −59.1 | −0.002 07 | 0.455 | −80.9 | −205 | −0.59 |
| $b_{477}$ | 0.215 | −0.0217 | 0.0231 | 252 | −0.043 | 0.003 34 | 0.049 |
| $b_{555}$ | −0.000 449 | −6.75 | 0.009 86 | −0.002 93 | 76.1 | −9.91 | 0.0205 |
| $b_{556}$ | 0.001 39 | 10.4 | −0.0263 | −0.0158 | −29.7 | 158 | −0.0479 |
| $b_{557}$ | 2.57 | −0.003 12 | 71.3 | −9.74 | 0.0581 | −0.0474 | 240 |
| $b_{566}$ | 0.005 79 | 23.2 | −0.0138 | 0.0161 | 158 | −0.405 | −0.0349 |
| $b_{567}$ | −17.9 | 0.118 | −19.6 | −81.1 | −0.0376 | 0.243 | −0.642 |
| $b_{577}$ | 0.004 84 | −7.5 | 0.006 27 | −0.0436 | 240 | −0.186 | −0.007 17 |
| $b_{666}$ | −0.0145 | 1.63 | −0.0311 | 0.139 | −0.136 | 357 | −0.105 |
| $b_{667}$ | 0.147 | −0.0237 | −12.7 | −103 | −0.0374 | −0.317 | 406 |
| $b_{677}$ | −0.005 76 | 36.6 | −0.0549 | 0.001 93 | −0.251 | 406 | 0.145 |
| $b_{777}$ | −0.0396 | −0.006 22 | 0.0474 | 0.0102 | −0.001 17 | 0.0459 | 295 |

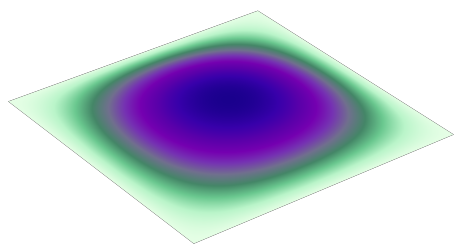
(a) 1ˢᵗ mode

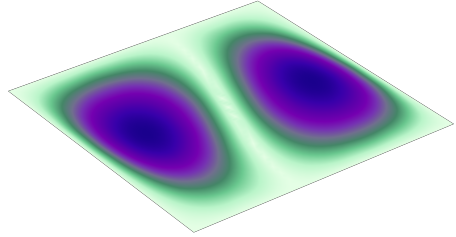
(b) 2ⁿᵈ mode

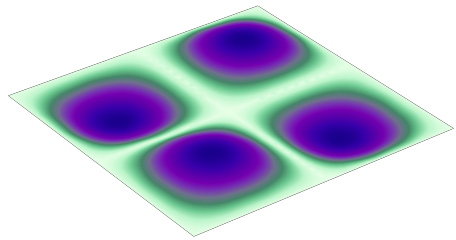
(c) 4ᵗʰ mode

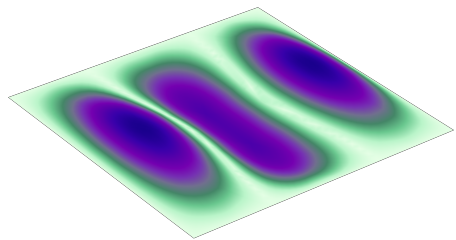
(d) 5ᵗʰ mode

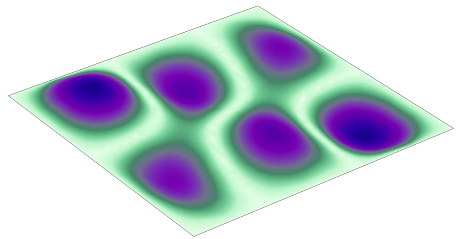
(e) 7ᵗʰ mode

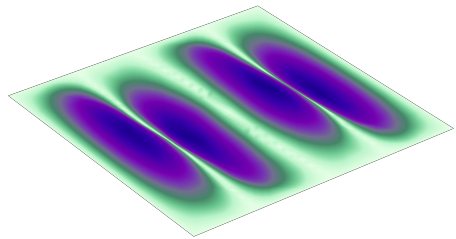
(f) 9ᵗʰ mode

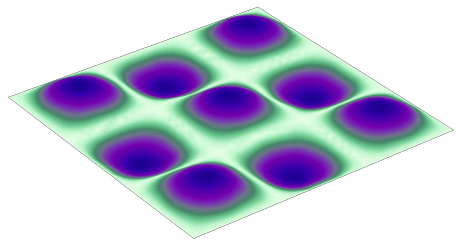
(g) 11ᵗʰ mode

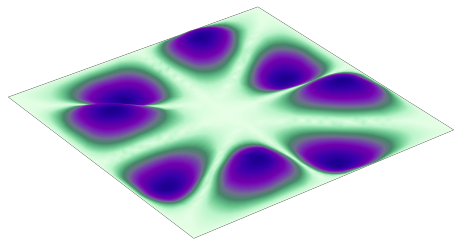
(h) 12ᵗʰ mode

Supplementary Figure 2: Eigenmodes of a square membrane resonator

## 2.3 Reduced-order model parameters for a string resonator

Reduced-order model of an ideal string resonator is based on an FE model constructed in COMSOL, considering pinned boundary condition with truss elements. The description of model parameters, scaling values and linear modal parameters can be found in Supplementary Tables 5, 9 and 10. We construct the model using the first three out-of-plane and first three in-plane modes of the structure. Nonlinear coefficients can be found in the Supplementary Tables 11 and 12.

Supplementary Table 9: Scaling values for a string resonator

|  | Linear | | | | | Nonlinear | |
|---|---|---|---|---|---|---|---|
| $m_\mathrm{t}$ | $k_{0_\mathrm{out}}$ | $k_{0_\mathrm{in}}$ | $\omega_{0_\mathrm{out}}$ | $\omega_{0_\mathrm{in}}$ | | $a_\mathrm{dim}$ | $b_\mathrm{dim}$ |
| $\rho A L$ | $\dfrac{\sigma A}{L}$ | $\dfrac{EA}{L}$ | $\dfrac{1}{2L}\sqrt{\dfrac{\sigma}{\rho}}$ | $\dfrac{1}{L}\sqrt{\dfrac{E}{\rho}}$ | | $\dfrac{EA}{L^2}$ | $\dfrac{EA}{L^3}$ |

Supplementary Table 10: Dimensionless linear modal values of a string resonator. Eigenfrequencies are normalized w.r.t. the fundamental mode ($\omega_1$). Note that Eq. 4, Eq. 5 & Eq. 6 describe in-plane modes.

|  | Eq. 1 | Eq. 2 | Eq. 3 | Eq. 4 | Eq. 5 | Eq. 6 |
|---|---|---|---|---|---|---|
| $m/m_t$ | 0.5 | 0.5 | 0.499 | 0.5 | 0.501 | 0.497 |
| $k/k_0$ | 4.93 | 19.7 | 44.4 | 4.97 | 19.9 | 44.6 |
| $\omega/\omega_1$ | 1 | 2 | 3 | 23.6 | 47.2 | 70.9 |

Supplementary Table 11: Quadratic nonlinear coefficients for a string resonator modelled with modes: first three out-of-plane modes (correspond to 1,2,3 in the table) and first three in-plane modes (correspond to 4,5,6 in the table)

|  | Eq. 1 | Eq. 2 | Eq. 3 | Eq. 4 | Eq. 5 | Eq. 6 |
|---|---|---|---|---|---|---|
| $a_{11}$ | 0 | 0 | 0 | 7.75 | $-0.000\,153$ | $-0.0014$ |
| $a_{12}$ | 0 | 0 | 0 | 0 | 0 | 0 |
| $a_{13}$ | 0 | 0 | 0 | 46.5 | $-93.1$ | $0.006\,46$ |
| $a_{14}$ | 15.5 | 0 | 46.5 | 0 | 0 | 0 |
| $a_{15}$ | $-0.000\,152$ | 0 | $-93.1$ | 0 | 0 | 0 |
| $a_{16}$ | $-0.001\,32$ | 0 | $0.008\,06$ | 0 | 0 | 0 |
| $a_{22}$ | 0 | 0 | 0 | 0 | $-62.1$ | 0 |
| $a_{23}$ | 0 | 0 | 0 | 0 | 0 | 0 |
| $a_{24}$ | 0 | 0 | 0 | 0 | 0 | 0 |
| $a_{25}$ | 0 | $-124$ | 0 | 0 | 0 | 0 |
| $a_{26}$ | 0 | 0 | 0 | 0 | 0 | 0 |
| $a_{33}$ | 0 | 0 | 0 | $0.000\,349$ | $-0.001\,73$ | $-209$ |
| $a_{34}$ | 46.5 | 0 | $-0.001\,16$ | 0 | 0 | 0 |
| $a_{35}$ | $-93.1$ | 0 | $-0.001\,73$ | 0 | 0 | 0 |
| $a_{36}$ | $-0.001\,08$ | 0 | $-418$ | 0 | 0 | 0 |
| $a_{44}$ | 0 | 0 | 0 | 0 | $-186$ | 0 |
| $a_{45}$ | 0 | 0 | 0 | $-372$ | 0 | $1.12 \times 10^3$ |
| $a_{46}$ | 0 | 0 | 0 | 0 | $1.12 \times 10^3$ | 0 |
| $a_{55}$ | 0 | 0 | 0 | 0 | 0 | 0 |
| $a_{56}$ | 0 | 0 | 0 | $1.12 \times 10^3$ | 0 | $0.0162$ |
| $a_{66}$ | 0 | 0 | 0 | 0 | $-0.0111$ | 0 |

Supplementary Table 12: Cubic nonlinear coefficients for a string resonator modelled with modes: first three out-of-plane modes (that correspond to 1, 2, 3 in the table) and first three in-plane modes (that correspond to 4, 5, 6 in the table)

| | Eq. 1 | Eq. 2 | Eq. 3 | Eq. 4 | Eq. 5 | Eq. 6 |
|---|---|---|---|---|---|---|
| $b_{111}$ | 18.3 | 0 | 18.3 | 0 | 0 | 0 |
| $b_{112}$ | 0 | 146 | 0 | 0 | 0 | 0 |
| $b_{113}$ | 54.8 | 0 | 328 | 0 | 0 | 0 |
| $b_{114}$ | 0 | 0 | 0 | 48.7 | −48.8 | 0.00533 |
| $b_{115}$ | 0 | 0 | 0 | −48.8 | 195 | 146 |
| $b_{116}$ | 0 | 0 | 0 | 0.00403 | 146 | 437 |
| $b_{122}$ | 146 | 0 | 219 | 0 | 0 | 0 |
| $b_{123}$ | 0 | 438 | 0 | 0 | 0 | 0 |
| $b_{124}$ | 0 | 0 | 0 | 0 | 0 | 0 |
| $b_{125}$ | 0 | 0 | 0 | 0 | 0 | 0 |
| $b_{126}$ | 0 | 0 | 0 | 0 | 0 | 0 |
| $b_{133}$ | 328 | 0 | −0.0227 | 0 | 0 | 0 |
| $b_{134}$ | 0 | 0 | 0 | 146 | −292 | −437 |
| $b_{135}$ | 0 | 0 | 0 | −292 | −0.00655 | 876 |
| $b_{136}$ | 0 | 0 | 0 | −437 | 876 | −0.0632 |
| $b_{144}$ | 48.7 | 0 | 73 | 0 | 0 | 0 |
| $b_{145}$ | −97.5 | 0 | −292 | 0 | 0 | 0 |
| $b_{146}$ | 0.00747 | 0 | −437 | 0 | 0 | 0 |
| $b_{155}$ | 195 | 0 | −0.00386 | 0 | 0 | 0 |
| $b_{156}$ | 292 | 0 | 876 | 0 | 0 | 0 |
| $b_{166}$ | 437 | 0 | −0.0141 | 0 | 0 | 0 |
| $b_{222}$ | 0 | 292 | 0 | 0 | 0 | 0 |
| $b_{223}$ | 219 | 0 | $1.31 \times 10^3$ | 0 | 0 | 0 |
| $b_{224}$ | 0 | 0 | 0 | 292 | 0 | −292 |
| $b_{225}$ | 0 | 0 | 0 | 0 | 781 | 0 |
| $b_{226}$ | 0 | 0 | 0 | −292 | 0 | $1.75 \times 10^3$ |
| $b_{233}$ | 0 | $1.31 \times 10^3$ | 0 | 0 | 0 | 0 |
| $b_{234}$ | 0 | 0 | 0 | 0 | 0 | 0 |
| $b_{235}$ | 0 | 0 | 0 | 0 | 0 | 0 |
| $b_{236}$ | 0 | 0 | 0 | 0 | 0 | 0 |
| $b_{244}$ | 0 | 292 | 0 | 0 | 0 | 0 |
| $b_{245}$ | 0 | 0 | 0 | 0 | 0 | 0 |
| $b_{246}$ | 0 | −583 | 0 | 0 | 0 | 0 |
| $b_{255}$ | 0 | 781 | 0 | 0 | 0 | 0 |
| $b_{256}$ | 0 | 0 | 0 | 0 | 0 | 0 |
| $b_{266}$ | 0 | $1.75 \times 10^3$ | 0 | 0 | 0 | 0 |
| $b_{333}$ | 0.00417 | 0 | $1.48 \times 10^3$ | 0 | 0 | 0 |
| $b_{334}$ | 0 | 0 | 0 | 438 | −439 | 0.0182 |
| $b_{335}$ | 0 | 0 | 0 | −439 | $1.76 \times 10^3$ | −0.0141 |
| $b_{336}$ | 0 | 0 | 0 | 0.00627 | 0.00105 | $3.93 \times 10^3$ |
| $b_{344}$ | 73 | 0 | 438 | 0 | 0 | 0 |
| $b_{345}$ | −292 | 0 | −877 | 0 | 0 | 0 |
| $b_{346}$ | −437 | 0 | 0.0373 | 0 | 0 | 0 |
| $b_{355}$ | 0.000582 | 0 | $1.76 \times 10^3$ | 0 | 0 | 0 |
| $b_{356}$ | 876 | 0 | −0.0161 | 0 | 0 | 0 |
| $b_{366}$ | −0.011 | 0 | $3.93 \times 10^3$ | 0 | 0 | 0 |
| $b_{444}$ | 0 | 0 | 0 | 292 | 0 | −290 |
| $b_{445}$ | 0 | 0 | 0 | 0 | $2.34 \times 10^3$ | 0 |
| $b_{446}$ | 0 | 0 | 0 | −875 | 0 | $5.24 \times 10^3$ |
| $b_{455}$ | 0 | 0 | 0 | $2.34 \times 10^3$ | 0 | $-3.51 \times 10^3$ |
| $b_{456}$ | 0 | 0 | 0 | 0 | $-7.02 \times 10^3$ | 0 |
| $b_{466}$ | 0 | 0 | 0 | $5.24 \times 10^3$ | 0 | 0.502 |
| $b_{555}$ | 0 | 0 | 0 | 0 | $4.7 \times 10^3$ | 0 |
| $b_{556}$ | 0 | 0 | 0 | $-3.51 \times 10^3$ | 0 | $2.1 \times 10^4$ |
| $b_{566}$ | 0 | 0 | 0 | 0 | $2.1 \times 10^4$ | 0 |
| $b_{666}$ | 0 | 0 | 0 | 0.132 | 0 | $2.35 \times 10^4$ |

# 3  Dynamic response of the graphene drum during internal resonance

In the main text, experiments were displayed where complex nonlinear phenomena occur, one of which is internal resonance caused by the interaction of the parametric resonance of the first mode with other modes of vibration. In the simulations, these observations were replicated. Here, we further investigate this response. In Supplementary Fig. 3, we show the time response of each mode during once cycle of the internal resonance at $\omega_d/\omega_0 = 2.538$. It is possible to observe strong activation of multiple modes. The dominant modes are the first three axisymmetric modes $f_{0,1}$, $f_{0,2}$ and $f_{0,3}$ (mode numbers 1,6,15) together with asymmetric mode $f_{2,1}$ (mode number 4). This is also clear in the overall frequency responses in Supplementary Fig. 4 with all the simulated modes.

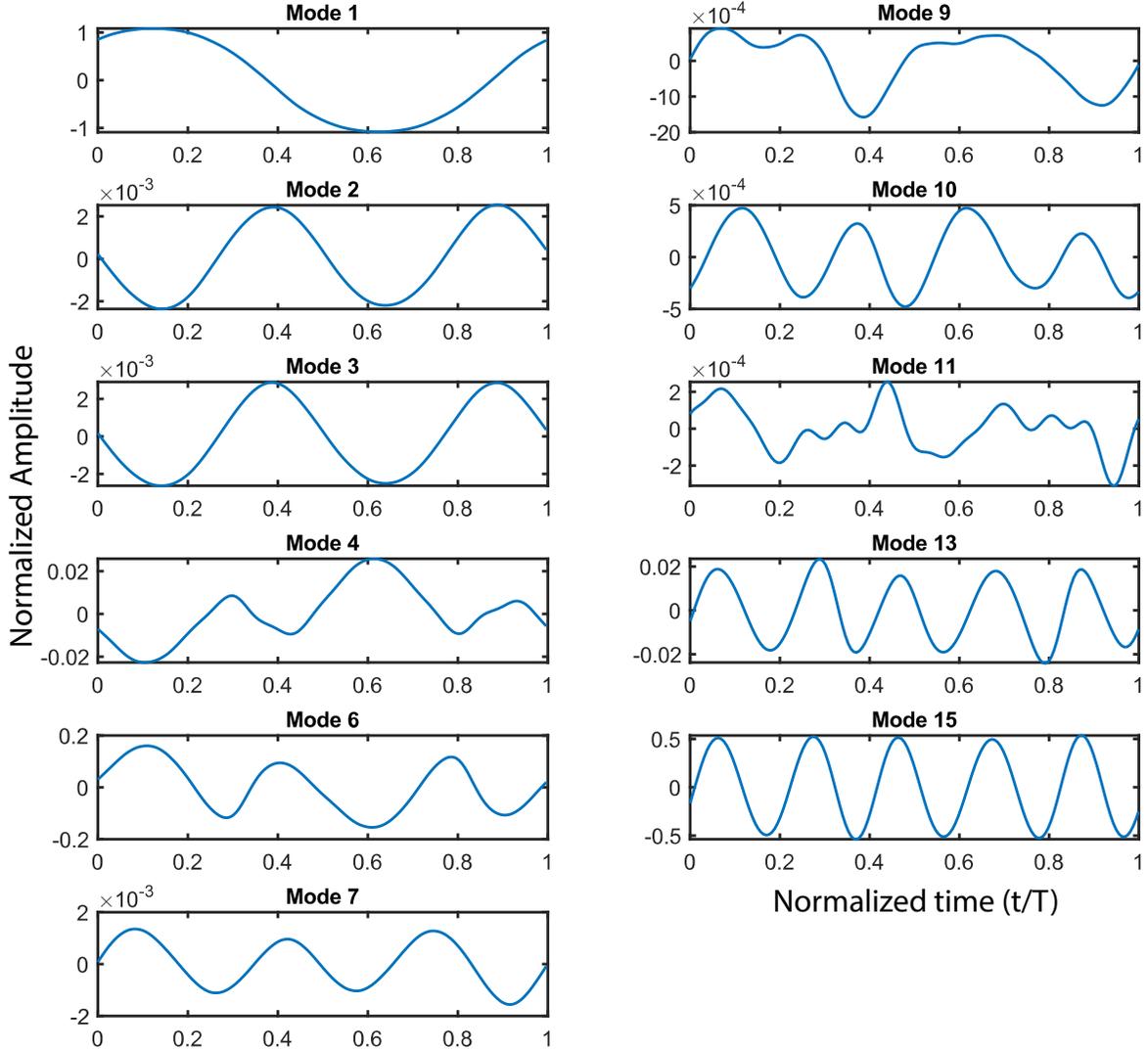

Supplementary Figure 3: Time responses of each mode during a period of oscillation for the graphene drum undergoing internal resonance at $\omega_d/\omega_0 = 2.538$.

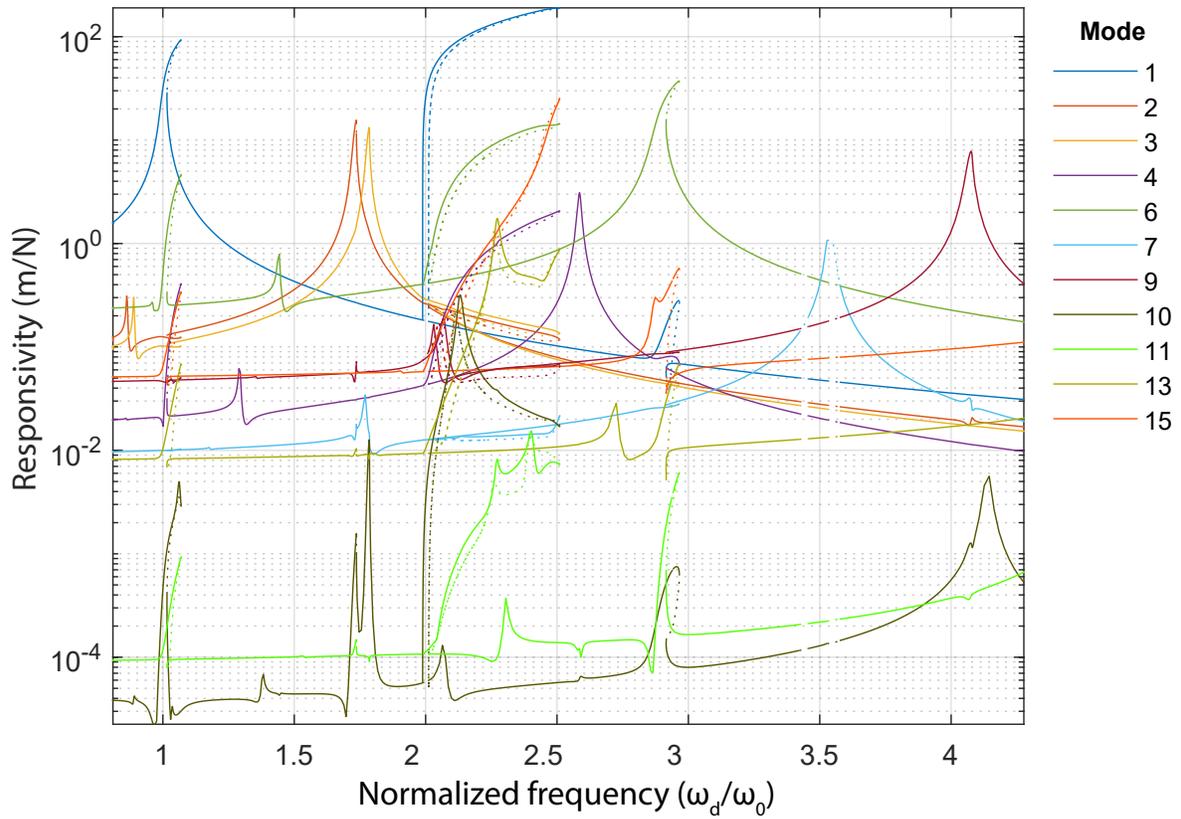

(a) Simulated modal frequency response for excitation level $\beta = 14.32$ and $\gamma = 2.5$, before the surge.

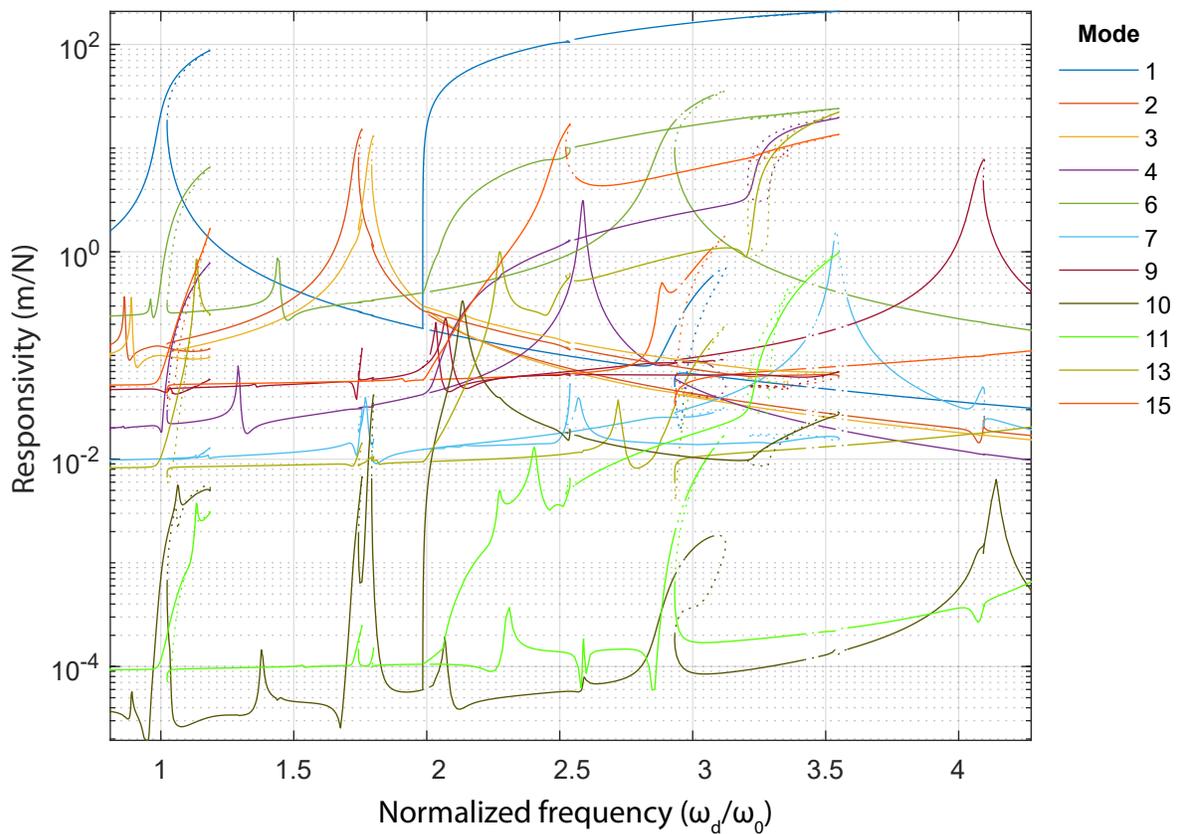

(b) Simulated modal frequency response for excitation level $\beta = 26.22$ and $\gamma = 3.2$, after the surge.

Supplementary Figure 4: Global multi-mode frequency responses of the simulated graphene drum

13

\AtBeginShipout\AtBeginShipoutDiscard
\end{document}